\documentclass[emulateapj]{emulateapj}
\usepackage{natbib}
\usepackage{graphicx}

\begin{document}
\title{Spitzer IRS Spectroscopy of the 10 Myr-old EF Cha Debris Disk: Evidence for 
Phyllosilicate-Rich Dust in the Terrestrial Zone}
\author{Thayne Currie\altaffilmark{1}, Carey M. Lisse\altaffilmark{2}, Aurora Sicilia-Aguilar\altaffilmark{3}, 
George H. Rieke\altaffilmark{4}, Kate Y. L. Su\altaffilmark{4}}
\altaffiltext{1}{NASA-Goddard Space Flight Center}
\altaffiltext{2}{Johns Hopkins University Applied Physics Laboratory, }
\altaffiltext{3}{MPIA-Heidelberg}
\altaffiltext{4}{Steward Observatory, University of Arizona}
\begin{abstract}
We describe \textit{Spitzer} IRS spectroscopic observations of the $\sim$ 10 Myr-old star, 
EF Cha.  Compositional modeling of the spectra from 5 $\mu m$ to 35 $\mu m$ confirms 
that it is surrounded by a luminous debris disk with L$_{D}$/L$_{\star}$ $\sim$ 10$^{-3}$, containing dust with  
temperatures between 225 K and 430 K characteristic of the terrestrial zone.
  The EF Cha spectrum shows evidence for many solid-state features, unlike most 
cold, low-luminosity debris disks but like some other 10--20 Myr-old luminous, warm debris disks (e.g. HD 113766A).  
The EF Cha debris disk is unusually rich in a species or combination of species whose emissivities 
resemble that of finely-powdered, laboratory-measured phyllosilicate species (talc, saponite, and smectite), which 
are likely produced by aqueous alteration of primordial anhydrous rocky materials.  
The dust and, by inference, the parent bodies of the debris also contain abundant amorphous silicates and metal sulfides, and possibly water ice.  
  The dust's total olivine to pyroxene ratio of $\sim$ 2 also provides evidence of aqueous alteration. 
The large mass volume of grains with sizes comparable to or below the radiation
blow-out limit implies that planetesimals may be colliding at a rate high enough to yield the emitting dust 
but not so high as to devolatize the planetesimals via impact processing.
Because phyllosilicates are produced by the interactions between 
anhydrous rock and warm, reactive water, EF Cha's disk is a likely signpost 
for water delivery to the terrestrial zone of a young planetary system.
\end{abstract}

\section{Introduction}
The 10--20 Myr age range is an important time in the formation of icy planetesimals and 
planets around $\sim$ 2 M$_{\odot}$ stars.
While debris emission decays with time for t $>$ 20 Myr as expected for simple collisional 
grinding of a fixed-mass planetesimal belt \citep{Rieke2005,Wyatt2007}, debris emission
 for A/F stars is observed to rise from 5 Myr to 10 Myr and peak from 
10--20 Myr \citep[][]{Currie2008,Hernandez2009,Cpk2008,CurrieLada2009}.
The rise in debris emission has been attributed to 
the agglomerate accretion of 1000 km-sized icy bodies in outer disk regions 
as planet formation switches from a fast runaway growth stage to 
a slower, debris-producing oligarchic stage \citep{KenyonBromley2008,KenyonBromley2010}.  
Dynamical perturbations by massive planets may also explain 
the luminous debris emission at these ages \citep{MustillWyatt2009,KennedyWyatt2010}.

\textit{Spitzer} has observed a number of luminous debris disks in this critical stage. 
A wide range of processes appear to be responsible for their extreme infrared emission, 
potentially revealing a variety of situations prevailing in their terrestrial planet zones.
For example, HD 113766A, an F3/5 star in 10-16 Myr-old Lower Centaurus Crux, shows 
evidence for multiple debris belts, including 
a warm (T $\sim$ 440 K) silicate dominated debris disk with a composition 
 consistent with S-type asteroids and whose flux densities require 
an enormous reservoir (M $\gtrsim$ M$_{Mars}$) of colliding material \citep{Chen2006, Lisse2008}.  \citet{Lisse2008} 
interpreted these spectral features as evidence of terrestrial 
planet-forming events.  
HD 172555, a $\sim$12 Myr-old A5 star, also exhibits luminous warm debris disk emission at 
L$_{D}$/L$_{*}$ $\sim$ 10$^{-3}$, but contains 
spectral features consistent with silica and SiO gas produced by a giant hypervelocity collision, similar 
to that likely responsible for forming the present-day Earth-Moon system \citep{Lisse2009}.
Eta Corvi, an ~1 Gyr old F2 star, shows an extended cold Kuiper Belt 
disk due to collisions driven by dynamical stirring, and a separate belt of warm
 inner system dust due to the collision of a KBO with a planet in the terrestrial habitable zone 
\citep{Wyatt2005,Lisse2010}.

Spatially-resolved images of other luminous debris disks at these ages  -- such as HR 4796A (A0V)
and $\beta$ Pic (A5V) -- find evidence for debris disk dynamical sculpting, such 
as asymmetries or warping, plausibly due to the presence of massive planets 
\citep{Mouillet1997,Golimowski2006,Schneider2009,Lagrange2010}.  However, 
unlike HD 113766A and HD 172555, they have rather featureless 
IR spectra, lacking evidence for massive collisions \citep{Chen2006}.
The disk around 49 Ceti (A0V) extends at high surface brightness to a radius of 900 AU 
but appears to have been cleared within 20 AU of the star, again possibly due to an unseen planet 
\citep{Wahhaj2007}.  Modeling of the images suggests that the mid-infrared emission from the 49 Ceti disk 
is dominated by small, short-lived grains ($\sim$ 0.01 $\mu m$) \citep{Wahhaj2007}.

EF Chameleonis (hereafter EF Cha) provides another interesting example of the variety of disk 
properties at this stage.  It is an $\sim$ 10 Myr-old A9IV/V star 95 pc from the Sun 
identified from the IRAS satellite as having infrared excess emission consistent 
with a luminous debris disk of L$_{D}$/L$_{\star}$ $\sim$ 10$^{-3}$ (for L$_{\star}$/ L$_{\odot}$ $\sim$ 9.7).  
A combination of MSX, ground based, and \textit{Spitzer} data show that its disk 
contains warm dust at temperatures $\sim$ 240 K \citep{Rhee2007}.  
It is among the relatively rare disks whose emission is dominated by warm dust and, therefore, 
by material in the terrestrial-planet zone \citep{Rhee2007,Currie2007a,Currie2007b,Morales2009}. 

In this paper, we present and analyze high signal-to-noise 5--35 $\mu m$ 
spectra of the EF Cha debris disk obtained with  
\textit{Spitzer}.  Section 2 describes 
our observations and data reduction.  In Section 3, 
we compare EF Cha to a simple model for blackbody 
disk emission and confirm that much of its IR excess is 
most plausibly due to debris emission originating 
from EF Cha's terrestrial planet-forming region.  
We then compare the EF Cha spectrum to spectra from 
other debris disks, protoplanetary disks, comets, and sophisticated 
mineralogical models in Section 4 to determine the chemical composition 
of the EF Cha debris disk dust and constrain the likely 
origin of its material.  We conclude by placing the spectral analysis of the EF Cha debris 
disk in the context of mineralogical studies of the early solar system and discuss 
processes likely responsible for explaining the EF Cha disk's spectrum.

\section{Observations and Data Reduction}
EF Cha was observed by the \textit{Spitzer Infrared Spectrograph} (IRS) on September 14 and 
October 12, 2008 in staring mode (PID 50150; AORs r25679360, r25678080, and r25678336).  
The low-resolution spectra covered 5--38 $\mu m$, 
using all four low-resolution orders (Short-Low 1 \& 2, Long-Low 1 \& 2), with exposures 
consisting of 2 cycles with a ramp duration of 6 seconds each. 
The high-resolution spectra covered 9.9--19.6 $\mu m$ in two cycles with a ramp duration of 30 seconds each. 
For the high-resolution spectra only, we obtained separate background observations.

We processed all spectra using the Spectroscopic Modeling Analysis and Reduction Tool 
\citep[SMART,][]{Higdon2004}.  
For the low-resolution spectra, we first median combined 
all frames for a specific setting (e.g. Short-Low 1) at a given nod position.  Then, 
we subtracted the median combined frame for one nod position from each frame corresponding to 
the other nod position, producing skysubtracted and bad pixel mitigated focal plane images.
  We then grouped the frames together by calibration file and extracted 
spectra from them simultaneously, using the automatic \textit{optimal} extraction algorithm
provided by SMART.  Finally, we grouped the extracted spectra for each setting together, 
masking the bonus order, clipping obvious outlying pixels and others using the 
SMART "special" clipping algorithm set at a 2.5 $\sigma$ threshold, masking pixels lying outside 
the wavelength range where IRS calibration is reliable (e.g. $\lambda$ $>$ 38 $\mu m$), and 
 median combined all spectra.  

For the high-resolution spectra, we first median-combined the science frames and the 
sky background frames at a given nod position.  The appropriate median-combined sky 
frame was then subtracted from the science frames at each nod position.  Using 
the \textit{full aperture} algorithm, we extracted each sky-subtracted frame.  
The spectra were then grouped together; obvious outlying pixels are identified by 
comparing pixel values from the four separate exposures, masked, and removed.  
Each spectrum was defringed with the algorithm \textit{irsfringe} ported into 
SMART and trimmed of pixels with unreliable flux calibration as determined 
from Table 5.1 in the IRS data handbook.  
As with the low-resolution spectra, we clipped the processed spectra at 2.5 $\sigma$ and 
 median combined them to obtain the final high-res spectrum.

\begin{figure}
\epsscale{0.8}
\centering
\plotone{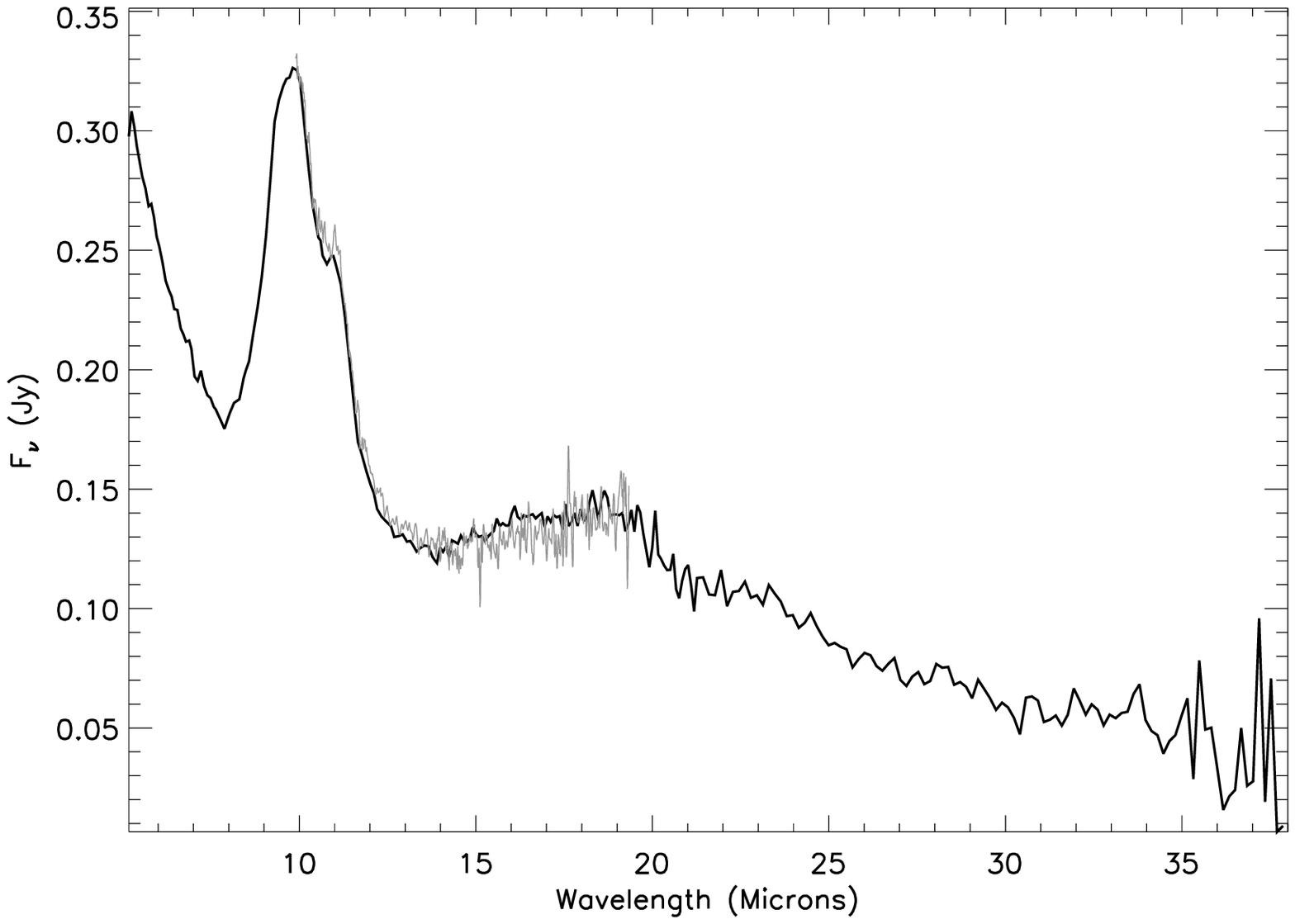}
\plotone{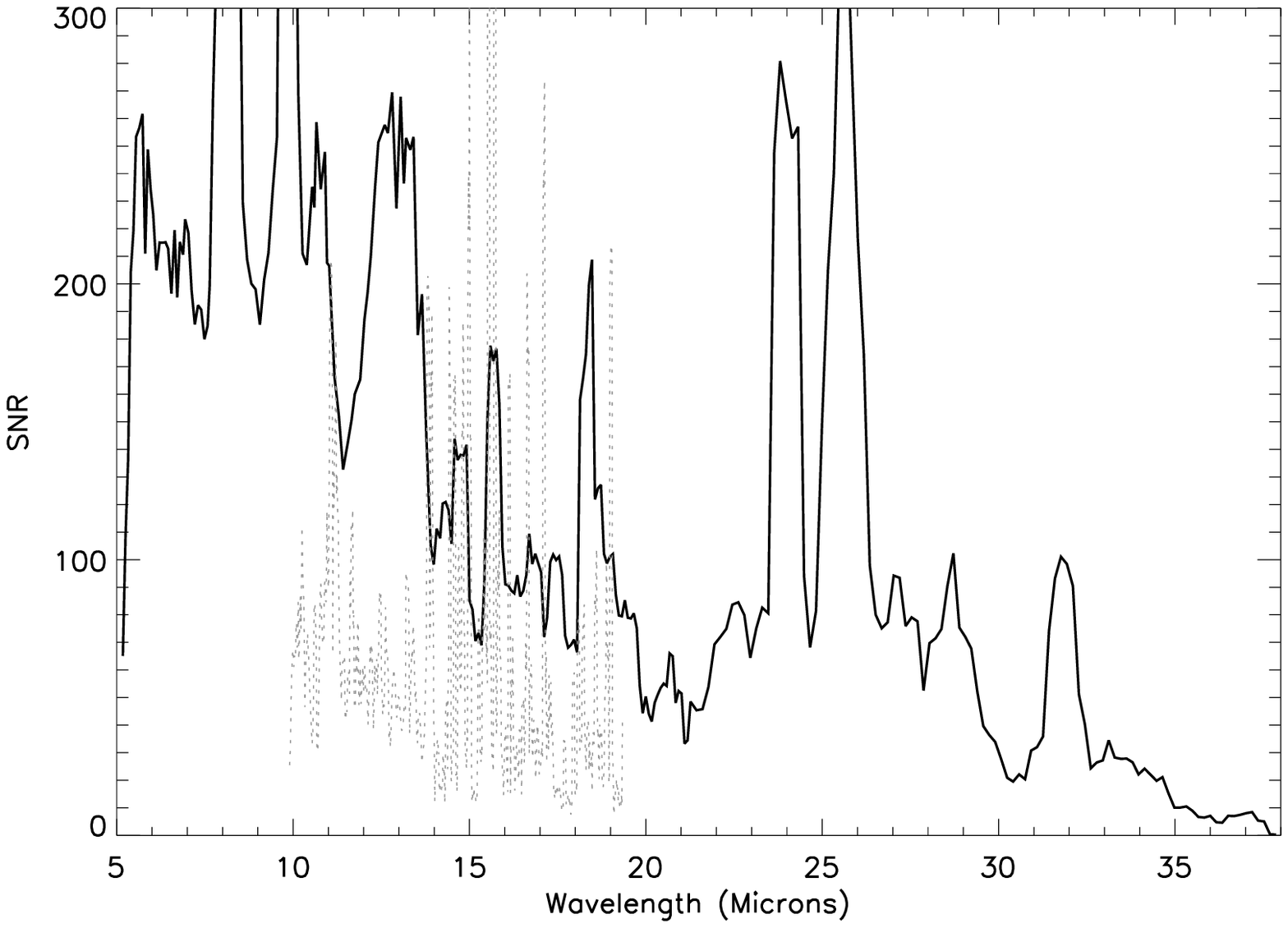}
\caption{(Top Panel) \textit{Spitzer} IRS low resolution (black line) 
 and high resolution (grey line) spectrum of EF Cha. (Bottom panel) Signal-to-noise 
of the low resolution and high-resolution spectra vs. wavelength.  The plotted 
signa-to-noise is smoothed by five pixels in wavelength.}
\label{efchaspec}
\end{figure}

Figure \ref{efchaspec} (top panel) shows the extracted spectra as a function of wavelength.  The low-resolution 
spectrum (black line) appears to be smooth and high quality from 5 through $\sim$ 30--35 $\mu m$.  
As expected, the high-resolution spectrum (grey line, binned by two wavelength elements) is 
noisier yet faithfully tracks the SED of the low-res spectrum.  
According to SMART, the low spectra achieve a signal-to-noise 
ratio $\sim$ 100-300 from 9 through 15 $\mu m$ and above 5--10 $\sigma$ at all 
wavelengths (bottom panel).  The signal-to-noise ratio for the high-res spectrum ranges from $\sim$ 50 
at 10 $\mu m$ to $\sim$ 10 at 19 $\mu m$.

\begin{figure}
\epsscale{1}
\plottwo{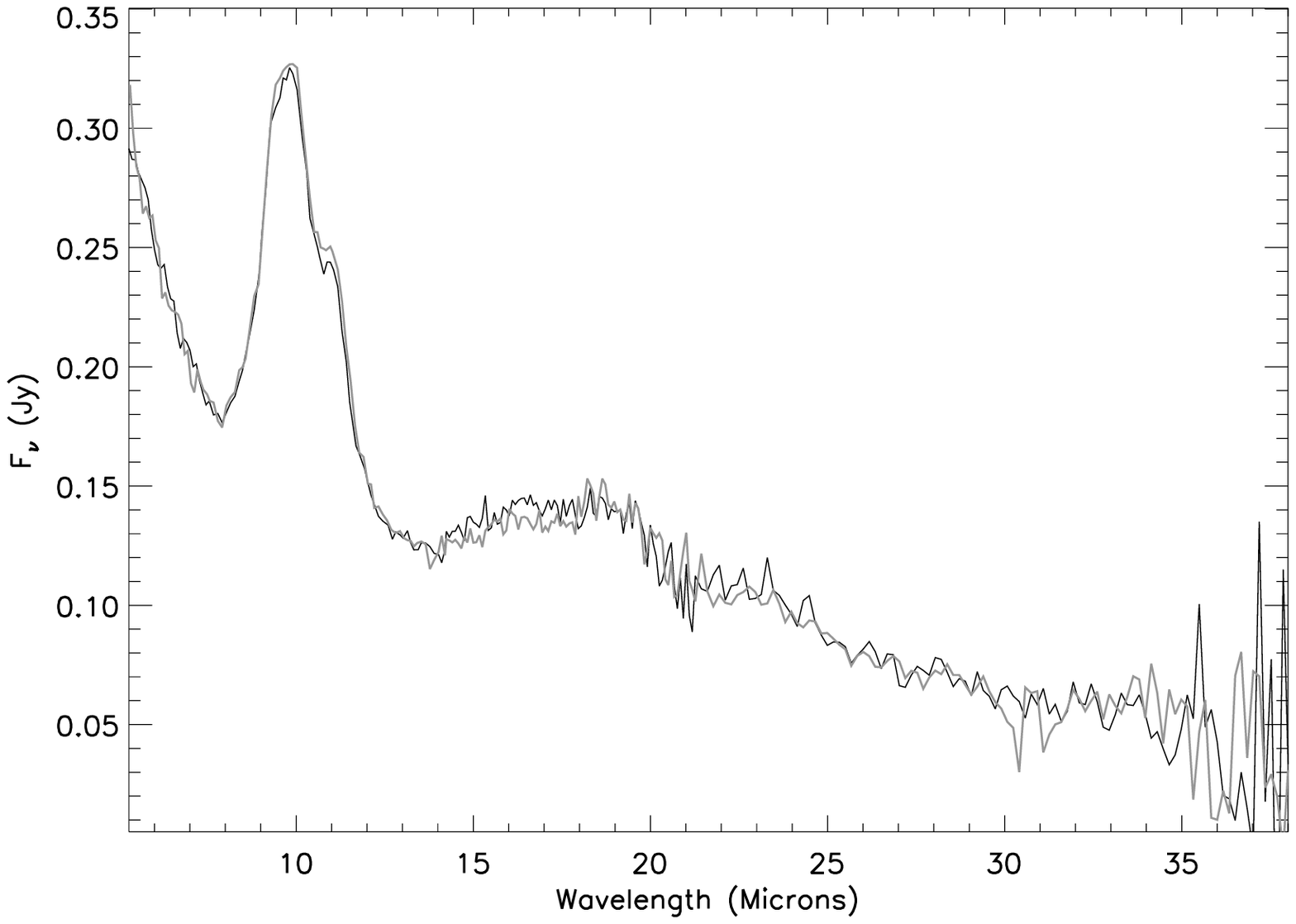}{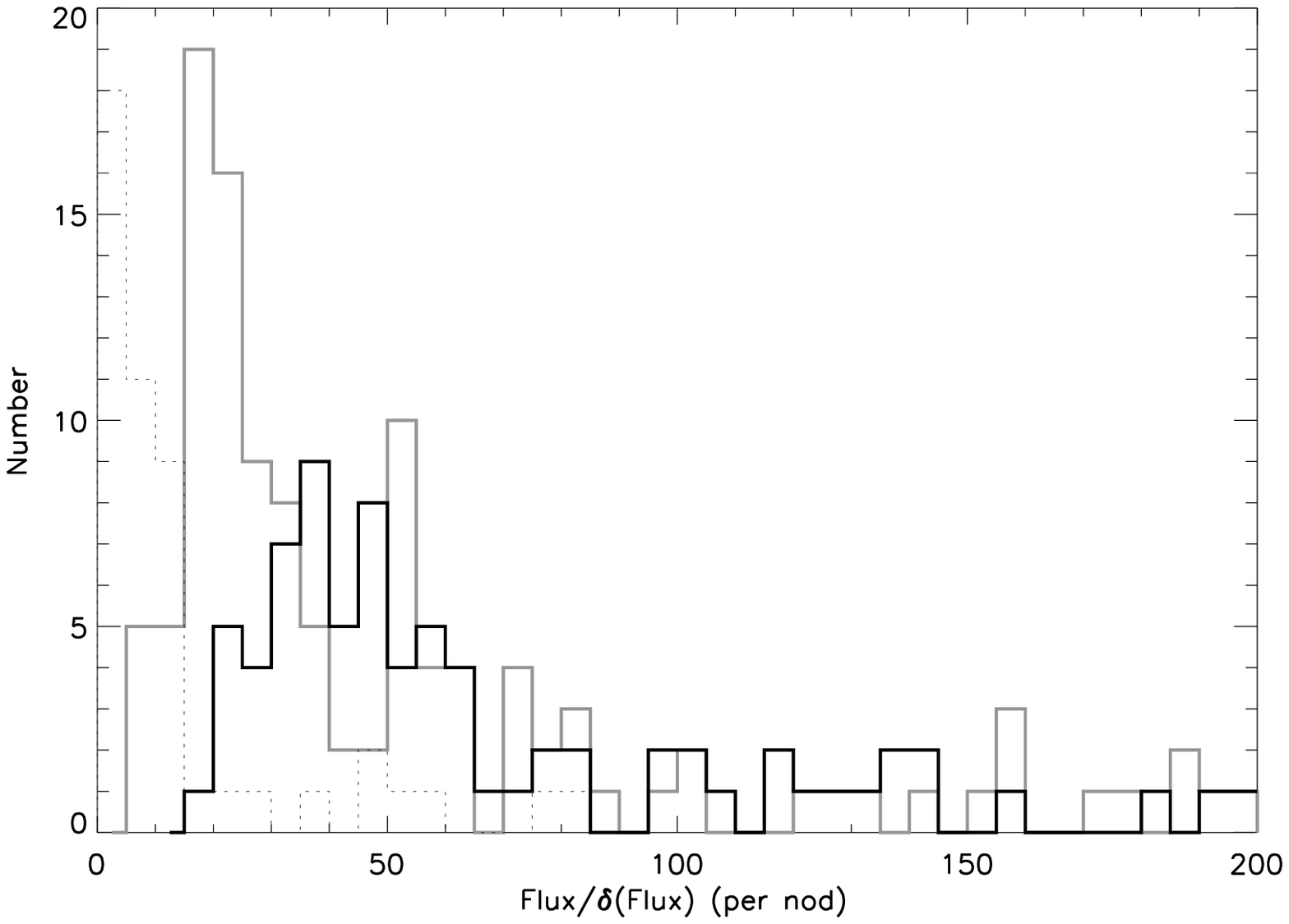}
\caption{(Left) Extracted low-res spectra for each of the two nod positions (black and grey lines).  (Right) 
Histogram plot of the flux of the first nod position divided by the flux difference between the nod positions.  
The distributions are separated by wavelength: 6--15 $\mu m$ (black line), 15--30 $\mu m$ (grey line), 
and 30--38 $\mu m$ (dotted line).}
\label{noddiff} 
\end{figure}

To provide a check on the signal-to-noise and reliability of our 
low-res spectrum, we performed two tests.  First, we extracted spectra from the 
two nod positions separately and then compared the flux densities at each wavelength 
(Figure \ref{noddiff}).  Through 15 $\mu m$ the spectra show very good agreement. 
The spectra become noisier at 15--30 $\mu m$ and then produce highly uncertain fluxes 
at $\sim$ 30--38 $\mu m$.  These comparisons allow us to empirically estimate the signal-to-noise per nod 
position as a function of wavelength (right panel).  Through 15 $\mu m$ the median signal-to-noise
is $\sim$ 57 per nod.
From 15 $\mu m$ to 30 $\mu m$, the median signal-to-noise per nod is $\sim$ 37.
 The median signal to noise at 30--38 $\mu m$ is 6.4.  
The median signal-to-noise of the combined and extracted spectrum should then be $\sim$ 81 through 15 $\mu m$, 
$\sim$ 52 from 15--30 $\mu m$, and $\sim$ 9 at 30--38 $\mu m$.  While these estimates are about 
a factor of two lower than those for the combined spectrum as derived by SMART, they confirm that 
the spectrum has a high signal-to-noise through $\sim$ 30 $\mu m$.

\begin{figure}
\epsscale{0.8}
\plotone{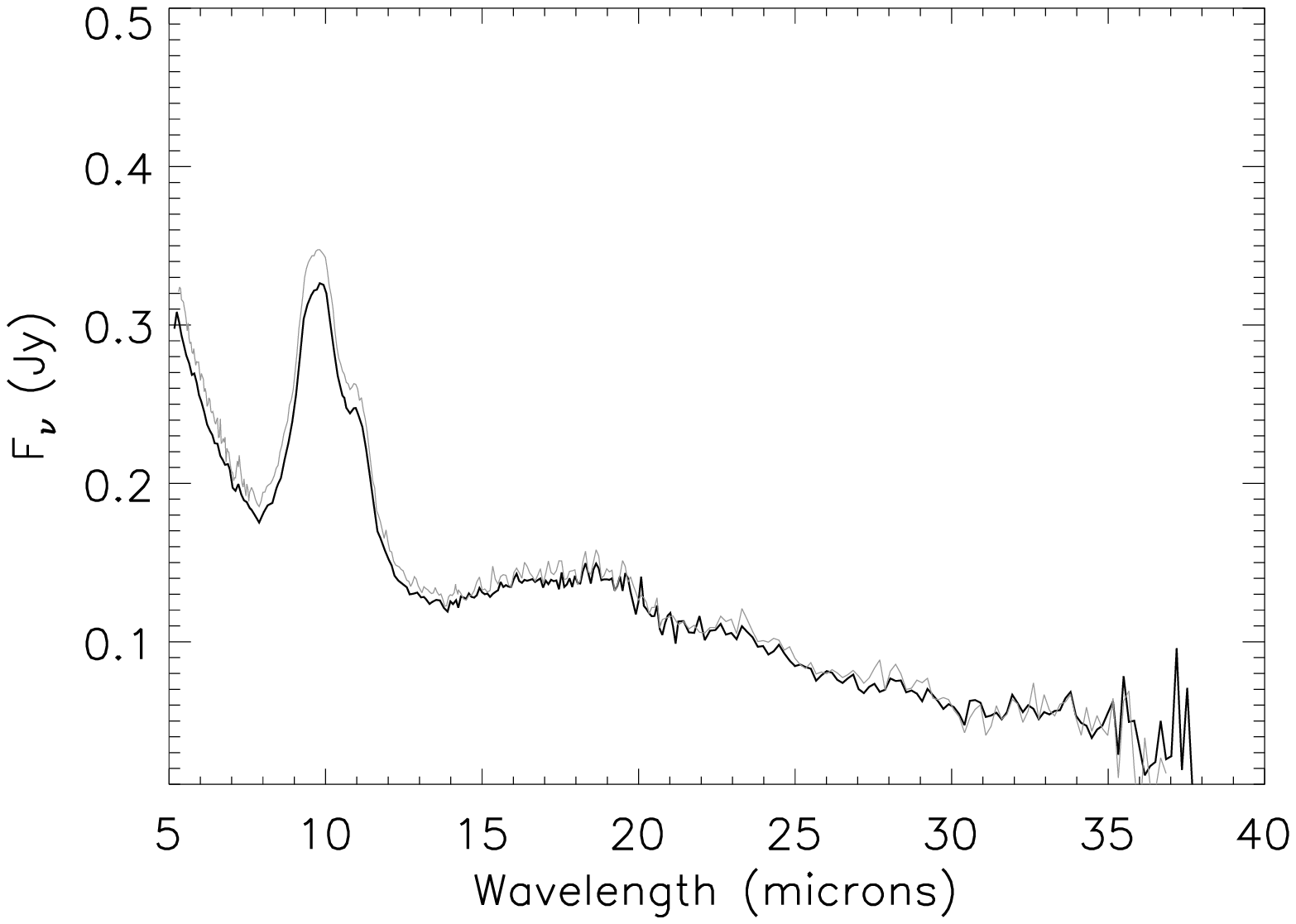}
\plotone{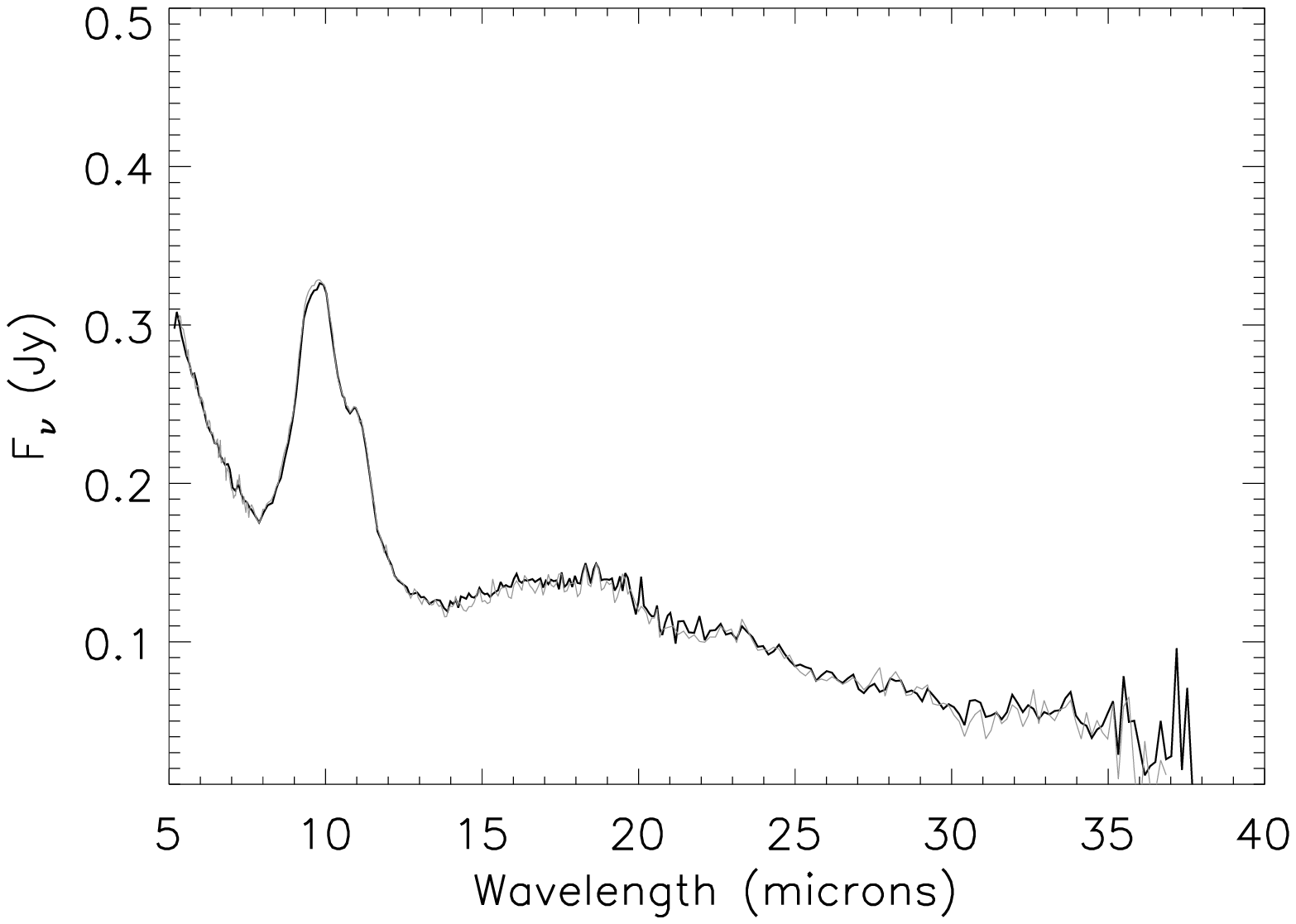}
\caption{Comparison between the EF Cha IRS spectral extraction produced from SMART and that produced from 
the FEPS pipeline (top panel).  (Bottom panel) The same comparison with the FEPS spectrum rescaled 
by 0.95, a 5\% decrease.}
\label{fepscomp}
\end{figure}

Second, we compared our SMART spectral extraction to that from the \textit{Formation and Evolution of 
Planetary Systems} (FEPS) pipeline \citep{Bouwman2008}.  Figure \ref{fepscomp} displays the two extractions.  
The FEPS pipeline-produced spectrum consistently has a higher flux than the SMART extraction (top panel).  
However, rescaling the FEPS spectrum by 0.95 yields an almost perfect agreement between the two 
extractions (bottom panel).  Since the FEPS and SMART-produced spectra are essentially scaled 
versions of one another, the differences likely arise from the way in which pointing offset 
refinements are treated, which compensate for the target not being precisely positioned 
at the center of the slit (J. Bouwman 2010, pvt. comm., \citealt{Swain2008}).  
  We conclude that our spectral extraction method is robust and interpreting the presence of 
spectral features, which depend on the \textit{relative} flux at different wavelengths,
 is not hindered by uncertainties/errors in the absolute flux calibration from our extraction method.

In part, these tests were performed to interpret strong 
features near $\sim$ 10 $\mu m$.  The IRS SL1 order has a permanent bad pixel lying very close to the 
spectrum at one of the nod positions near $\sim$ 10 $\mu m$: 
failure to properly identify and mitigate this pixel can lead to 
a spurious strong emission peak (G. Bryden 2009, pvt. comm.).  Given the excellent 
agreement between spectra extracted from 
different nod positions and that extracted using the FEPS pipeline, 
we conclude that we have succesfully corrected for any such bad pixel-induced spectral artifacts.


\section{Analysis}

\subsection{Basic Properties of the EF Cha Disk: Disk Evolutionary State and Dust Temperature}

First, we examine the spectra, looking for any evidence of nebular gas emission expected 
if the EF Cha disk is a "remnant" protoplanetary disk instead of a debris disk.
In neither the low nor the high-resolution spectra identify evidence for H$_{2}$ gas emission at 
12.1 $\mu m$, 17 $\mu m$, or 28 $\mu m$ expected for disks with copious amounts of 
gas \citep[e.g.][]{Gorti2004,Pascucci2006, Carmona2008,MartinZaidi2010}.  Neither do we find evidence for [Ne II] 
emission at $\sim$ 12.8 $\mu m$, indicative of trace amounts ($\sim$ 10$^{-6}$ M$_{J}$) 
of circumstellar gas \citep{Pascucci2007}.  
 There may be a very faint, low signal-to-noise 
emission feature at $\sim$ 6 $\mu m$ corresponding to water gas, though its significance 
is less than 3 $\sigma$.  Since this emission, if real, can 
be produced by evaporating icy bodies it does not imply the presence of substantial 
primordial nebular gas.  

Detecting gas emission lines like H$_{2}$ and [Ne II] has proved 
difficult for younger stars with masses like EF Cha's \citep[e.g.][]{Juhasz2010}, 
some of which like AB Aur have direct evidence for circumstellar gas \citep[][]{Bitner2007}.  
Because clear mid-IR detections of gas may only be possible for small fractions of the 
Herbig AeBe disk lifetime \citep[e.g.][]{MartinZaidi2010}, the absence of 
these lines does not guarantee the absence of gas in the system.
  However, less than $\sim$ 1--2\% of 
stars as old as EF Cha show evidence for gas \citep[e.g.][]{Currie2007c}, 
and most protoplanetary disks around M $\approx$ 2 M$_{\odot}$ stars 
disappear prior to $\sim$ 3 Myr \citep[][]{CurrieKenyon2009}.
The EF Cha disk is also optically thin at all wavelengths \citep{Rhee2007} and its emission 
can be well fit with one or two single temperature blackblodies consistent with 
narrow debris belts \citep[][Sections 3.2 and 3.3 below]{Rhee2007}.
Protoplanetary disks, on the other hand, are typically optically thick in the mid-IR 
and are better fit by disks with large spatial extents.
Both the low resolution and high-resolution spectra contain a strong, broad 
10 $\mu m$ silicate emission feature that is due to the presence of many 
small, micron-sized dust grains.  Thus, to first order the mid-IR spectrum 
of EF Cha is consistent with that expected for a gas-poor, dusty debris disk, 
as argued by \citet{Rhee2007}.

\begin{figure}
\plotone{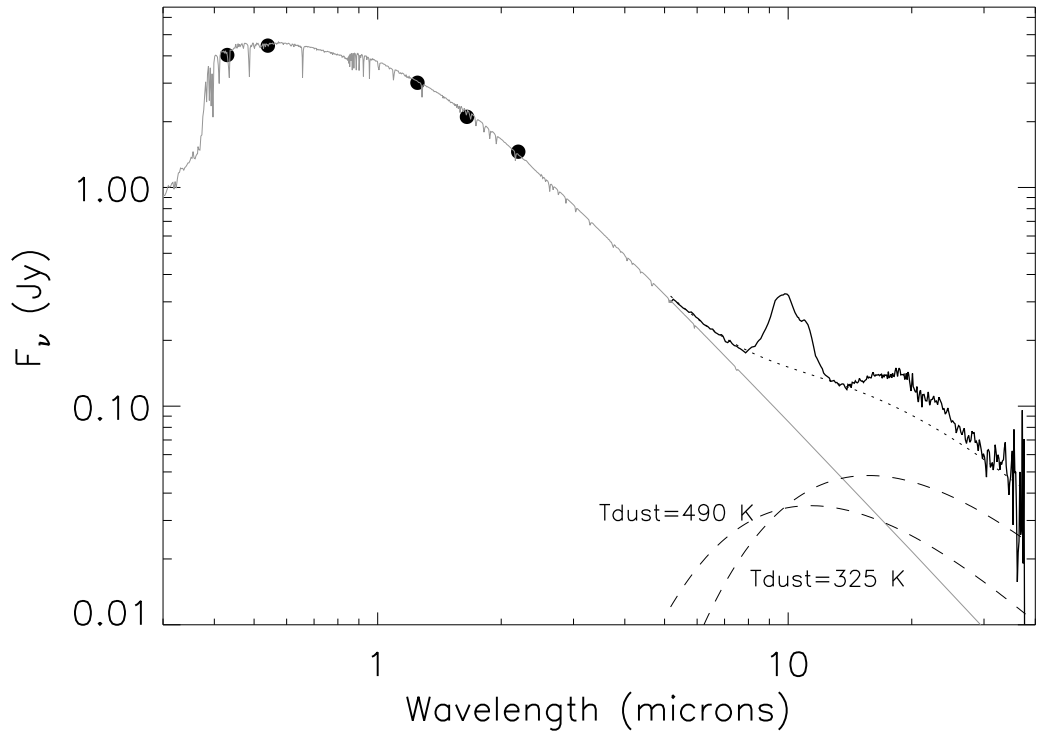}
\caption{EF Cha SED including photometric data from the Tycho-II survey and 2MASS (dots) and IRS data (solid black line).  
The dashed lines correspond to the star + single-temperature blackbodies; the dotted line corresponds to the combined 
flux from these blackbodies+the star's flux.  For the star's flux, we fit the SED to a Kurucz stellar atmosphere model for a T$_{e}$ $\sim$ 7400 K star (grey line).}
\label{efchased}
\end{figure}

Next, we construct a full SED of the EF Cha system to determine the temperature of the dust producing the observed IRS excess emission. 
To our \textit{Spitzer} data, we add B and V data from the Tycho-II catalog and 
JHK$_{s}$ data from the 2MASS All-Sky Survey \citep[][Figure \ref{efchased}]{Skrutskie2006}.
We fit the optical data to the appropriate Kurucz stellar atmosphere model for an A9 star, using the effective temperature scale 
from \citet{Currie2010a}.  Removing the stellar photosphere model from the total IRS spectrum, we find disk excess emission rising above the noise level at $\sim$ 5--6 $\mu m$ and dominating the system emission by 10 $\mu m$.

To make an initial determination of the typical dust temperature, we fit the observed 
SED to a simple two-temperature blackbody model, with dust temperature ranges 
of 50-400 K for the cool component and 300-1000 K for the hot component.  In 
 doing this, we required that the model reproduce the observed SED in regions 
lacking obvious solid state features ($\lambda$ $\sim$ 9--12 $\mu m$, 15--25 $\mu m$).  
The SED is best modeled by dust belts with temperatures of $\sim$ 325 K and 490 K, corresponding to dust in the terrestrial 
zone of the system.  
More generally, the SED at 30--35 $\mu m$ is consistent with a F$_{\nu}$ $\propto$ $\lambda^{-2}$ decline expected for 
warm dust in the Rayleigh-Jeans limit.  Assuming a luminosity of 9.7 L$_{\odot}$ for EF Cha and the 
blackbody relations between dust temperature and location -- T$_{dust}$ $\sim$ 280 
K$\times$(r/1 AU)$^{-1/2}$(L$_{star}$/L$_{Sun}$)$^{1/4}$
 these temperatures imply the presence of dust belts at 1.0 AU and 
2.4 AU.  As we were able to fit the entirety of the observed flux to within 20\%, we conclude, based on these simple fits,  
that EF Cha lacks evidence for any cold debris component.

Since the EF Cha disk spectrum does not resemble a featureless 
blackbody as in most debris disks \citep[e.g.][]{Chen2006}, accurately determining dust temperatures 
is complicated by the presence of solid state features.  Moreover, accounting for these features may shift the best-fit 
dust temperatures and thus the inferred location of the dust.  Nevertheless, we can conclude that EF Cha's disk emission 
is consistent with emerging from one or two debris belts at temperatures higher than the 0 torr water-ice condensation 
temperature ($\sim$ 170 K; $\sim$ 8.5 AU from the star at LTE).  Thus, EF Cha shows evidence for a warm debris disk with emission extending from 
the terrestrial zone to regions analogous to the solar system's asteroid belt, confirming earlier work 
by \citet{Rhee2007}.

\subsection{Comparisons with Protoplanetary Disk Spectra and Typical Debris Disk Spectra: Empirical Limits on Dust Composition}
The mid-IR spectrum of the EF Cha debris disk shows evidence for 
many solid state features.  To investigate the composition of the dust disk, 
we first compare its emission to that from protoplanetary disks 
around Herbig AeBe stars and typical debris disks around 1--3 M$_{\odot}$ stars.
  Herbig AeBe star spectra exhibit a wide 
range of mid-IR spectral features, ranging from the comet-like spectrum of HD 100546 
 to the 10 $\mu m$ silicate-free spectrum of HD 169142 \citep{Meeus2001,Grady2005}.

\begin{figure}
\centering
\includegraphics[scale=0.45]{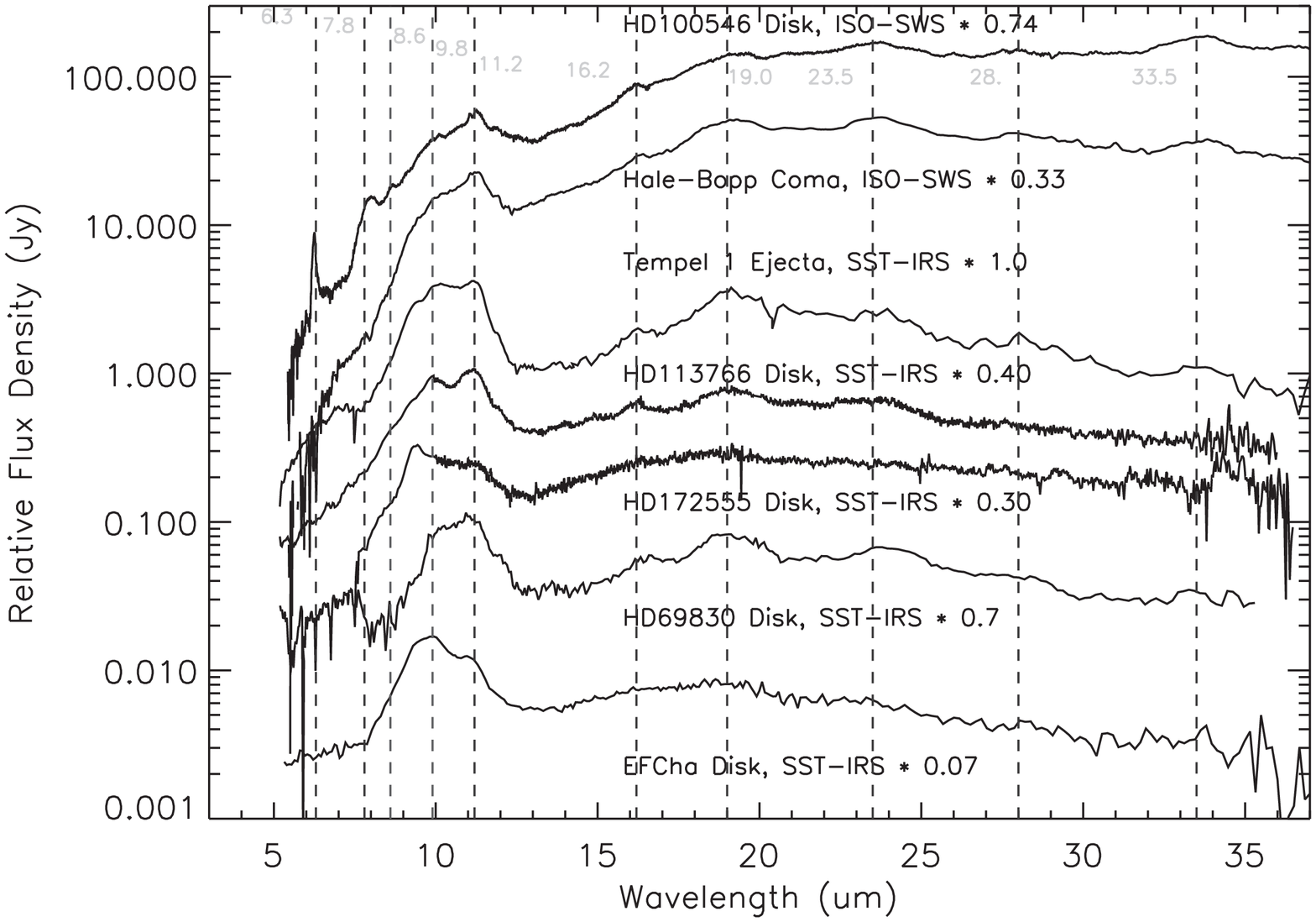}
\\
\includegraphics[scale=0.45]{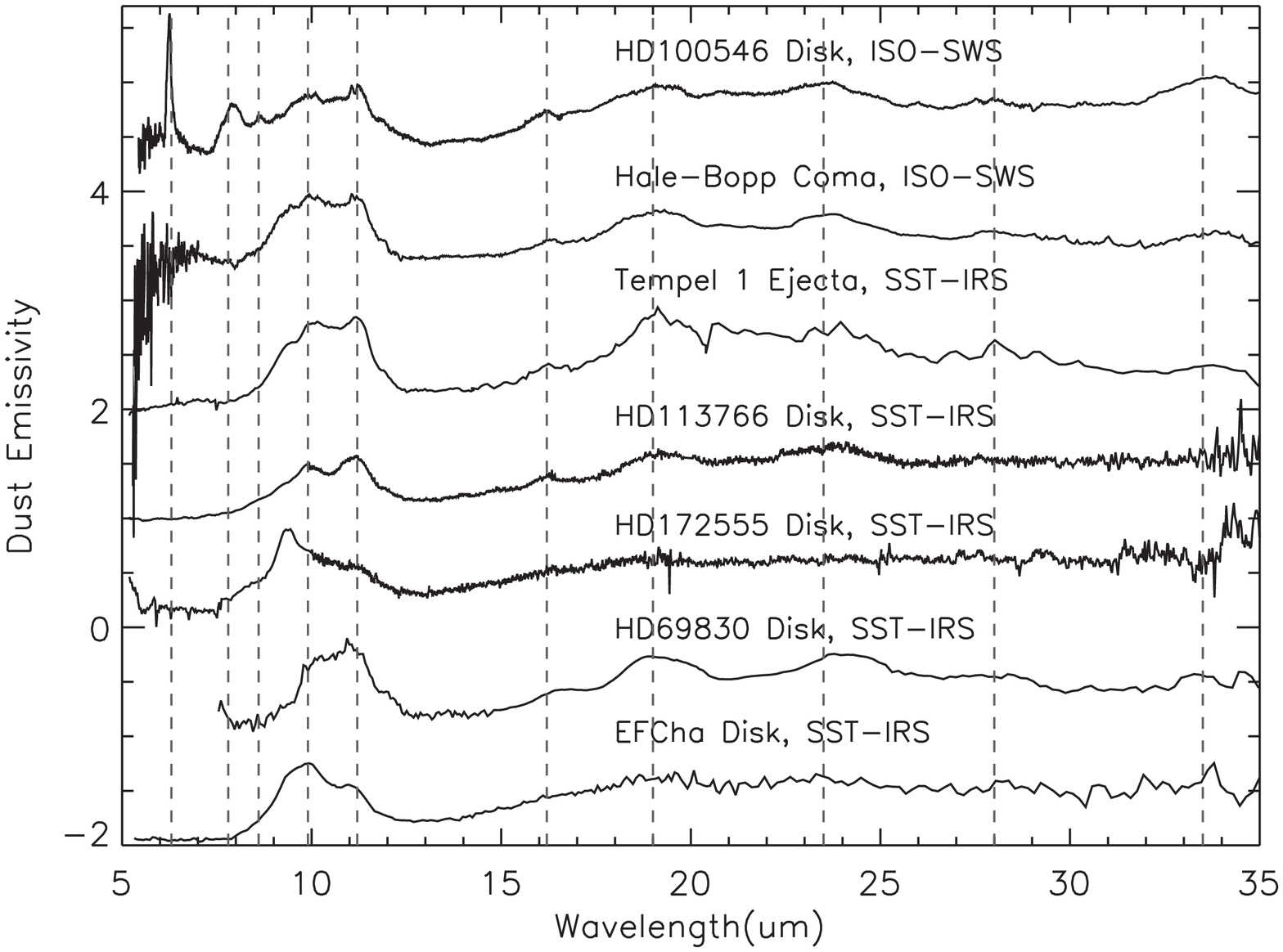}
\caption{(Top) Spectra of a protoplanetary disk (HD 100546), Comet Hale-Bopp, the 
ejecta from Comet Tempel-1, and several high-luminosity debris disks -- HD 113766, 
HD 172555, and HD 69830 compared to that for EF Cha \citep{Lisse2008,Lisse2009,Lisse2007b}.  Wavelengths for PAH emission and other solid state 
features are identified by vertical dashed lines. 
(Bottom) Spectra redisplayed in emissivity space to better contrast the feature to continuum flux ratios for each object.}
\label{fluxcompare}
\end{figure}

The EF Cha spectrum is significantly different than that of the Herbig AeBe star HD 100546 (Figure 
\ref{fluxcompare}) primarily in that it lacks strong PAH emission features at 6.3 $\mu m$, 7.7 
$\mu m$, and 8.6 $\mu m$.  
Compared to other protoplanetary disk spectra studied by \citet{Meeus2001}, EF Cha's disk 
shares the greatest similarity with disks like HD 150193 and HD 163296 labeled as "Group IIa", which have broad, "boxy" 
solid state features at $\sim$ 10 microns and generally lack clear PAH emission features
at 6.2, 7.7, 8.6, and 11.2 $\mu m$.  EF Cha shows a secondary bump at $\sim$ 11.2 $\mu m$ --
comparable in wavelength to a PAH emission feature such as that exhibited by HD 100546 but also easily explained 
by emission from olivine grains -- lacks PAH emission features at shorter wavelengths.  

Most debris disk spectra are featureless; for example, even when the dust is 
warm enough to emit significantly at 10 $\mu m$, debris disks usually 
lack the intrinsically strong 10 $\mu m$ silicate feature \citep[e.g.][]{Dahm2009,
Morales2009}.  The lack of silicate features indicates that the dust grains are large 
 with diameters $\ge$ 10 $\mu m$ \citep[e.g.][]{Chen2006}.  
In sharp contrast, the EF Cha spectrum shows a complex 10 $\mu m$ peak and 
spectral structure longwards of $\approx$ 15 $\mu m$, indicating that the disk contains 
many grains smaller than 10 $\mu m$.   This behavior is qualitatively similar to 
that of other warm, high-luminosity debris disks around young stars, such as those 
found around HD 113766A and HD 172555 \citep[][Figure \ref{fluxcompare}]{Lisse2008,Lisse2009}.

High mass fractions of crystalline 
pyroxene can also be ruled out since they have multiple, narrow emission peaks in the 8--13 $\mu m$ 
range resulting in a spectrum far less smooth than observed.  Likewise, the spectrum 
cannot contain as high of an abundance of crystalline olivine as that exhibited 
by comets and some Herbigs like the HD 100546 disk: 
such a high abundance would result in a narrow maximum at $\sim$ 9.7 $\mu m$ and a 
higher maximum at $\sim$ 11.2 $\mu m$ \citep[see ][]{Lisse2007}, the 
opposite of what is observed.  The general
shape of the EF Cha spectrum is instead consistent with a high abundance of amorphous silicates 
and a secondary population of olivine.
Amorphous silicates produce  
much broader 8--13 $\mu m$ peaks, resulting in a secondary maxima at $\sim$ 17--19 $\mu m$, 
and are abundant in Group IIa objects \citep[e.g.][]{Bouwman2001,Lisse2007}.  
However, superimposed upon this shape are peaks at 10, 11, 19, 23, 28, and 
33 $\mu m$.  The existence of these peaks indicates that additional species 
besides amorphous silicates are present.

\subsection{Comparisons with Mineralogical Models}
\subsubsection{Description of Mineralogical Model}
As shown by the empirical comparisons, the composition of mid-IR 
emitting dust from EF Cha's debris disk is dominated by amorphous silicates 
but must include high mass fractions of other species.
To identify these additional species, we fit the EF Cha IRS spectrum with
a detailed mineralogical model based on laboratory thermal infrared emission 
spectra described in detail by \citet{Lisse2006,Lisse2007, Lisse2007b,Lisse2008}.
Below, we reiterate the key components of the model and describe its 
application to our data.

\textbf{Thermal Dust Emission} -- 
The emission from distribution of dust is given by 
\begin{equation}
F_{\lambda, model} = \frac{1}{\Delta^{2}}\sum_{i,a}\int_0^\infty 
B_{\lambda}(T_{i}(a,r_{\star}))Q_{abs,i}(a,\lambda) \pi a^{2} \frac{dn_{i}(r_{\star})}{da} da
\end{equation}
where T is the particle temperature for a particle of radius \textit{a} 
and composition \textit{i} at a distance \textit{r} from the central.  
the distance from \textit{Spitzer} to the dust is $\Delta$. B$_{\lambda}$ is 
the blackbody radiance at wavelength $\lambda$, Q$_{abs}$ is the 
emission efficiency of the particle of composition \textit{i} at 
wavelength $\lambda$, dn/da is the differential particle 
size distribution of the dust.  The sum is over all species and
all particle sizes.  Our spectral analysis consists of calculating 
the flux for a model distribution of dust and comparing 
the result to the observed flux.  The predicted flux from 
the model depends on the 1) dust composition (affects the location 
of spectral features), 2) particle sizes (affects feature-to-continuum 
contrast), and 3) the particle temperature (affects strength of 
short versus long wavelength features).   

\textbf{Dust Composition}-- We use thermal laboratory emission spectra  
drawn from measurements for randomly oriented, $\mu m$-sized powders, which 
yield Q$_{abs}$ for each species.  Sources for the emission spectra are 
detailed in the Supplemental Online Material from \citet{Lisse2006} and among 
others include contributions from the Jena spectral library (http://www.astro.uni-jena.de/Laboratory/OCDB) 
and from \citet{Koike2000,Koike2002,Chihara2002} for silicates.  The library of material spectra 
were selected by their reported presence in interplanetary dust particles, 
meteorites, in situ comet measurements, protoplanetary disks, and 
debris disks \citep[see][]{Lisse2006}.  

In total, the library  
 includes spectra from 80 different species.  
The list of materials considered include multiple silicates in the 
olivine and pyroxene class (forsterite, fayalite, clino- and 
ortho-enstatite, augite, anorthite, bronzite, diopside, and ferrosilite).  
The model also includes phyllosilicates (such as saponite, serpentine, 
smectite, montmorillonite, and chlorite); sulfates (such as gypsum, 
ferrosulfate, and magnesium sulfate); oxides (including various 
aluminas, spinels, hibonite, magnetite, and hematite) 
Mg/Fe sulfides (including pyrrohtite, troilite, pyrite, and 
niningerite); carbonate minerals (including calcite, aragonite, 
dolomite, magnesite, and siderite); water-ice, clean and with 
carbon dioxide, carbon monoxide, methane, and ammonia 
clathrates; carbon dioxide ice; graphitic and amorphous 
carbon; and neutral and ionized polyaromatic hydrocarbons (PAHs).

\textbf{Dust Size Effects} -- We consider a dust size range of 
0.1 to 1000 $\mu m$ in the model fits, 
with sharp emission features arising mainly  from particles of 0.1 - 10 $\mu m$, and  
baseline continuum mainly from 10 - 1000 $\mu m$ particles.
The particle size effects the dust emissivity as
\begin{equation}
1-Emissivity(a,\lambda)=[1-Emissivity(1 \mu m,\lambda]^{a/1 \mu m}
\end{equation}
The particle size distribution (PSD) is fit at log steps in radius, i.e., 
at 0.1, 0.2, 0.5, 1, 2, 5, ...10, 20, 50 $\mu m$, etc.  Particles of 
the smallest sizes have emission spectra with very sharp features, and 
little continuum emission.  Particles of larger sizes -- greater than $\sim$ 10 $\mu m$ -- 
tend to be optically thick, and contribute only featureless continuum emission.  
Particles larger than 1000 $\mu m$ are not directly identifiable from our modeling and 
contribute no emission.  We assume the same particle size distribution 
for each species.

\textbf{Dust Temperature} -- Dust particle temperatures are determined at the 
same log steps in grain radius used to determine the PSD.  Dust temperature 
is a function of dust composition, size, and astrocentric distance.  The highest 
temperature for the smallest particle size of each species is varied freely 
and is determined by the best fit to the data.  The largest, optically thick 
particles (1000 $\mu m$) are set to the LTE temperature, and the temperature of 
particles of intermediate sizes is interpolated between these two extremes 
by radiative energy balance.   

\textbf{Accuracy/Reliability of the Model and Fitting Method} --
This model has been successfully applied to both solar system objects and protoplanetary/debris 
disks around other stars. In particular, the model accounts for 
the emission spectra of comet 9P/Tempel 1 observed during the Deep Impact mission \citep{Ahearn2005, 
Lisse2006}. Prior to the collision with the spacecraft, the comet's IR spectrum was near featureless, 
consistent with a coma dominated by large blackbody grains, similar to debris disk spectra around most 
stars. After collision, the spectrum showed more than 16 pronounced emission features,
strong evidence for $\approx$ 10$^{7}$ kg of small, submicron to micron-sized grains consisting of amorphous and 
crystalline silicates, amorphous carbon, carbonates, phyllosilicates, PAHs, water gas and ice, 
and sulfides \citep{Lisse2006}. As evidenced by the copious presence of fragile, volatile water 
ice grains in the ejecta, the differences in the pre- and post-impact spectrum must be 
due to simple de-aggregation of weak, large ($\approx$ 100 $\mu m$ to 1 cm), optically thick 
fractal dust particles in the delivered shock wave, rather than mineralogical 
transformation of the dust at high temperatures and pressures due to the delivered 
energy. 

Direct compositional measurements from cometary 
sources also provide strong evidence that our modeling approach 1) is accurate for dust in 
the solar system and 2) can infer the composition of dust around other stars.  
A comparison of the derived Tempel 1 material composition with the materials 
found in the STARDUST sample return show good agreement (Hanner and Zolensky 2010). 
Moreover, the direct measurements of species within interplanetary dust particles 
(IDPs) of cometary origin, in particular the GEMS (glass with embedded metal and 
sulfides) grains located within the matrices of ``chondritic porous" IDPs, 
provide good evidence that cometary/meteoritic provide a good spectral match to 
``astronomical" silicates such as those inferred from spectra of protoplanetary 
and debris disks \citep{Bradley1999}.

For both Tempel 1 and debris disks around other stars like EF Cha, the dust responsible 
for mid-IR solid state features is produced from collisions and must be 
freshly produced as it is small and easily removed from the system by radiation pressure, 
Poynting-Robertson drag, and stellar wind drag.
Thus, applying our model to debris spectra from other stars can 1) 
 identify the chemical composition of parent bodies from which the debris originate and 
2) determine the sizes of debris particles, which may identify whether the debris must
originate from large scale, catastrophic impacts and 3) determine the range of dust 
temperatures.  \citet{Lisse2008,Lisse2009} argue that debris emission from disks surrounding HD 113766A and HD 172555 
originates from debris created by large-scale collisional grinding/giant impacts associated with terrestrial 
planet formation and/or stochastic events such as that responsible for forming the Moon.
Likewise, we here apply the model to the EF Cha spectrum to constrain 
the composition of parent bodies and investigate the origin of the debris.

Our method differs from other widely-used methods 
in that it uses physically plausible emission measures from randomly-oriented powders 
rather than theoretically-derived Mie values.   The free parameters of the model are the 
relative abundance of each detected mineral species, the temperature of the smallest 
dust particle of each species, and the value of the particle size distribution.  
Best-fits are found by a direct search through phase space -- composition, temperature, 
and size distribution.

\subsubsection{Results of Fits From Our Mineralogical Model}
\begin{figure}
\plotone{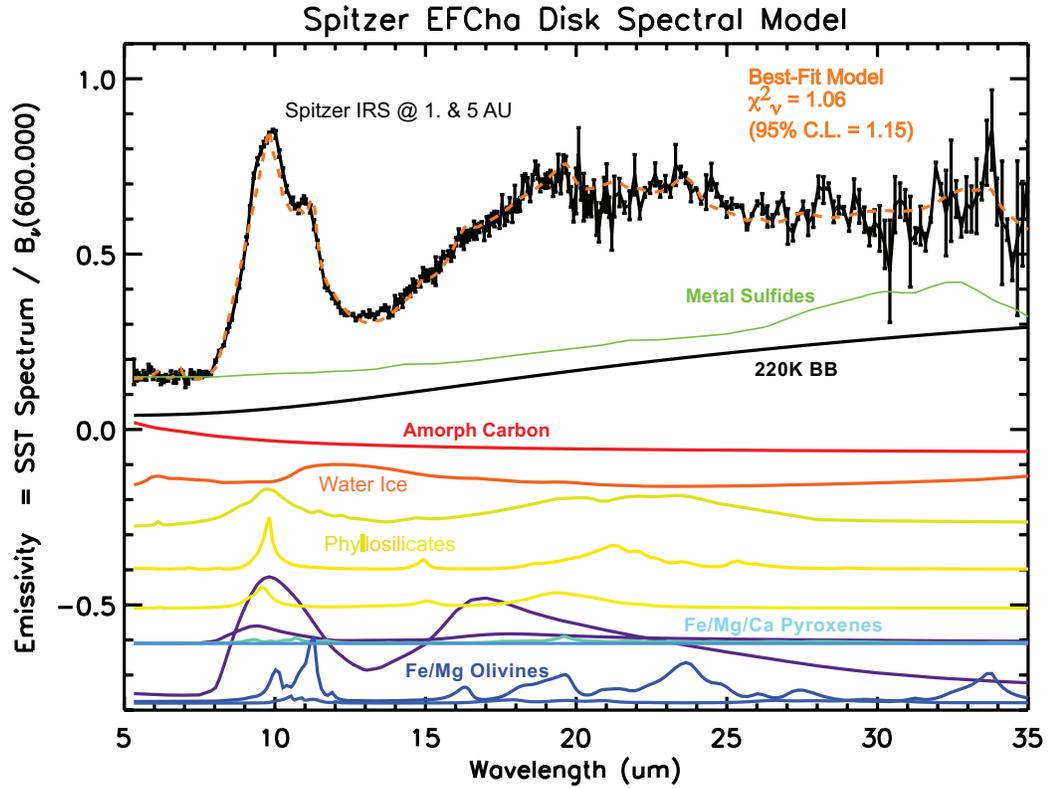}
\caption{Spitzer IRS emissivity spectrum of EF Cha with the best-fit spectral decomposition.  
The central source's photospheric contribution has been removed using a Kurucz model with a 7400 K 
photospheric temperature. Error bars are $\pm$ 2$\sigma$. The amplitude of each colored curve 
denotes the relative amount of that species present in the best-fit model (Table 1). For 
species with no statistically detectable emission, the curve is a flat horizontal line. 
Black points: \textit{Spitzer} dust excess spectrum, divided by a 600 K blackbody. Orange dashed line: 
best-fit model spectrum. Colored curves: Purples - amorphous silicates of pyroxene or 
olivine composition. Light blues - crystalline pyroxenes: ferrosilite, diopside, and orthoenstatite. 
Dark blues - crystalline olivine forsterites. Red - amorphous carbon. Deep orange - water ice. 
Light orange - water gas. Yellow lines- phyllosilicates. Olive green - ferromagnesian sulfide (Fe$_{0.9}$Mg$_{0.1}$S).
The peak at $\approx$ 10 $\mu m$ is caused by a combination of amorphous silicates and phyllosilicates; 
phyllosilicate species are responsible for structure at $\sim$ 15 $\mu m$, which is more readily apparent 
in the following figure.  Spectral structure at 16--25 $\mu m$ is due mostly to forsterite.}
\label{efchafit}
\end{figure}

Figure \ref{efchafit} shows our fit to the EF Cha flux, presented in emissivity space (spectrum divided 
by an average Planck function of the entire dust population) for clarity, with the emissivity of 
individual constituent species labeled and shown as colored lines. 
Table 1 lists the relative abundances of each species.
The best-fit model requires the presence of two warm dust belts, 
one with micron-sized dust at 
600 K (LTE at $\sim$ 430K, or $\sim$ 1.3 AU) and one with large dust 
grains ($>$ 20 $\mu m$) at an LTE temperature of 
$\sim$ 220 K ($\sim$ 5 AU).  

The best-fit model successfully reproduces the entire IRS spectrum and yields a small
  $\chi^{2}$ per degree of freedom ($\chi^{2}$/dof = 1.06).  To provide a statistical measure 
of the goodness-of-fit, we determine the 95\% confidence limit given the number of 
degrees of freedom for a given reduced $\chi^{2}$ \citep[see][]{Lisse2008,Lisse2009}.  The total number of free parameters -- 
including the number of relative abundances, number of hottest particle temperatures, and power law index -- 
is 32.  Given the number of spectral points fit (272), the number of degrees of freedom is 240.  
Based on these values, there is a 95\% chance a model with 
 reduced $\chi^{2}$ less than 1.15 is a good predictor of the Spitzer data.
Given the excellent fit of the 
model from 5 to 38 $\mu m$, it is plausible that the chemical abundances included 
in the model comprise the dominant detectable species in the disk. 

As suggested by the empirical comparisons, the EF Cha spectrum is olivine rich, containing 
large relative molar abundances of amorphous olivine and forsterite (N$_{moles, rel}$ = 0.73 and 
0.46, respectively) responsible for the large width of the 10 $\mu m$ feature and 
the presence of a secondary peak at $\sim$ 11 $\mu m$.  
Fayalite also is present in low abundance.  The model $\chi^{2}$$_{\nu}$ is 1.07 with this species removed, 
lower than the 95\% confidence limit, so evidence for this species is very weak. 
The emissivity slope and features at $\sim$ 20--35 $\mu m$ indicate the presence of 
water ice and metal sulfides, which are also abundant in 
the EF Cha disk (N$_{moles, rel}$ = 0.44 and 1.45, respectively)).  The spectrum contains 
a small abundance of pyroxene but lacks any 
clear evidence for water gas, carbonate, or PAH emission typical of primitive cometary material.  

\begin{figure}
\plotone{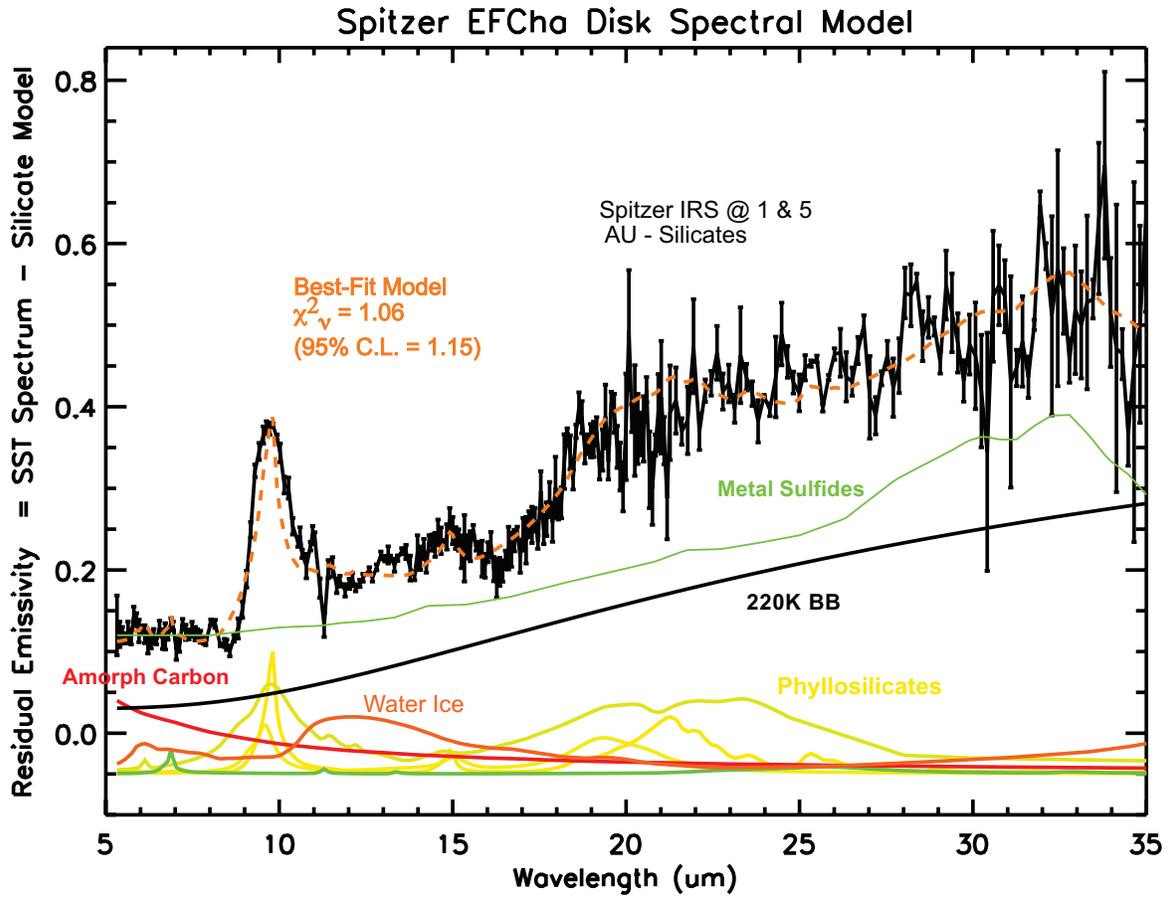}
\caption{Residual emissivity of the circumstellar dust, after the strong emission 
due to silicates has been fit and removed. The remnant is dominated by emission from phyllosilicates, 
amorphous carbon, water ice, and metal sulfides. 
While there are interesting hints of potential water gas emission at 6 $\mu m$, and 
of carbon dioxide gas emission at 15 $\mu m$, neither of these features is statistically significant. 
Emission peaks at 9.8 $\mu m$ and 15 $\mu m$ due to phyllosilicate species are visible.
}
\label{efchafitnosil}
\end{figure}

Interestingly, the EF Cha debris disk appears to contain a very high abundance 
of phyllosilicate species - smectite, talc, and saponite --
 responsible for the amplitude and shape of the pronounced peak 
at 10 $\mu m$,  15 $\mu m$ and 20--23 $\mu m$ (N$_{moles, rel}$ = 0.2 total).
Figure \ref{efchafitnosil}, displaying 
disk emissivity with silicate contributions subtracted out, more clearly shows 
the importance of phyllosilicates and other major constituents like metal sulfides.  
In particular, phyllosilicates have singular sharp features and are required to reproduce the 
strong residual 10 $\mu m$ emission and complicated slope of the emissivity at 17--25 $\mu m$.  
 SiO and silica have peaks at 8.8 to 9.2 $\mu m$ wavelengths, much shorter than 10 $\mu m$, 
and their inclusion in models clearly does not improve the $\chi^{2}$ parameter.
Without the presence of phyllosilicate species, the $\chi^{2}$$_{\nu}$ of the 
spectral fit is far worse -  3.1 for the removal of smectite alone.  
The range of abundances from any model passing the 95\% confidence 
limit is narrow, typically $\sim$ 10\% from the best-fit value.


Given the numerous strong solid state features, the EF Cha disk must contain large numbers of small 
grains.  The best-fit model includes a minimum grain size of 0.1 $\mu m$: two orders of magnitude 
smaller than grain sizes for most IRS-studied debris disks.  Moreover, the 
particle size distribution for the best-fit model is dn/da $\propto$ a$^{-4.0}$.  
This steep distribution implies the presence of many more fine particles than 
for a collisional equilibrium size distribution (dn/da $\propto$ a$^{-3.5}$).

While we can only directly sense the presence of grains with sizes 
$\lesssim$ 20 $\mu m$, grains up to millimeters in size have been found 
in debris disks \citep{Maness2008}.  Moreover, micron-sized dust results 
from the collisional cascade induced from colliding, km-sized planetesimals 
and thus implies the presence of much larger bodies \citep[e.g.][]{KenyonBromley2008,Wyatt2008,Krivov2010}.  
The size distribution of these fragments are predicted from theory to follow 
a power law behavior roughly continuous with that for directly 
detectable debris dust \citep[e.g. see][]{KenyonBromley2008}.
Therefore, we can get a rough estimate of the total mass of fragments involved in the collisions 
by extending the particle size distribution up to larger sizes of order of 1 mm.

Assuming a maximum grain size of 1000 $\mu m$, we estimate a total dust mass 
of $\sim$ 1.5$\times$10$^{23}$ g, comparable to the mass of an asteroid with 
a bulk density of 2.5 g cm$^{-3}$ and radius of $\sim$ 110 km.
This parent size is a lower limit since it is based on the total mass of 
grains with sizes less than 1 mm.  Larger, 100-1000 km-sized bodies are likely 
present since they are required to initiate and sustain the collisional 
cascade that produces the $<$ 1 mm-sized debris \citep[e.g.][]{KenyonBromley2008,KenyonBromley2010}.

The large number of small grains must require a replenishment source since 
they would otherwise be rapidly removed by radiation pressure 
\citep[e.g.][]{BackmanParesce1993, Chen2006}.  Assuming 
a stellar mass of 1.9 M$_{\odot}$, a stellar luminosity of 9.7 L$_{\odot}$, 
and a grain density of 3 g cm$^{-3}$, the blowout size limit 
for spherical grains in EF Cha, using the equation in \citet{CurrieKenyon2009}
 is $\sim$ 2 $\mu m$.  Thus, the EF Cha debris disk contains a high mass of dust dominated by 
particles comparable to or smaller than those dominating typical debris disks and also 
far smaller than the nominal blowout size limit for the system.
Along with the stabilty of the 1983 - 2007 IRAS/MSX/Spitzer photometry, the copious 
amount of small dust indicates that there must be a source for small particle replenishment.

\subsubsection{Tests of Phyllosilicate Identification}
Additional tests are required to more reliably constrain the presence of individual species 
responsible for the EF Cha debris emission, especially phyllosilicate species.  
First, although our best-fit model includes phyllosilicate species our solution 
is not unique if a phyllosilicate-free model reproduces the observed spectrum within 
the confidence limit we adopt to identify good-fitting models (the 95\% confidence limit).  
Second, the signatures from phyllosilicate species and other species like metal sulfides 
are weaker than the amorphous silicates also present in the disk.   Since they are weaker 
their presence is more likely to be falsely identified in real spectra with random noise patterns.  

To address these issues, we carried out two additional sets of model fits.
  First, we reran the fits with all phyllosilicate species removed, in order to 
determine whether all good-fitting models require their presence.  Second, 
we fit our models to two independent EF Cha spectra obtained concurrently at separate IRS 
nod positions.  Comparing the two model results constrains the sensitivity of our identification 
of different species to photon noise.
Moreover, since the spectra at different nod positions are extracted 
from different regions on the IRS detector, fits to separate nod positions 
test the sensitivity of our identifications to different IRS systematic noise contributions.

\begin{figure}
\includegraphics[scale=0.6]{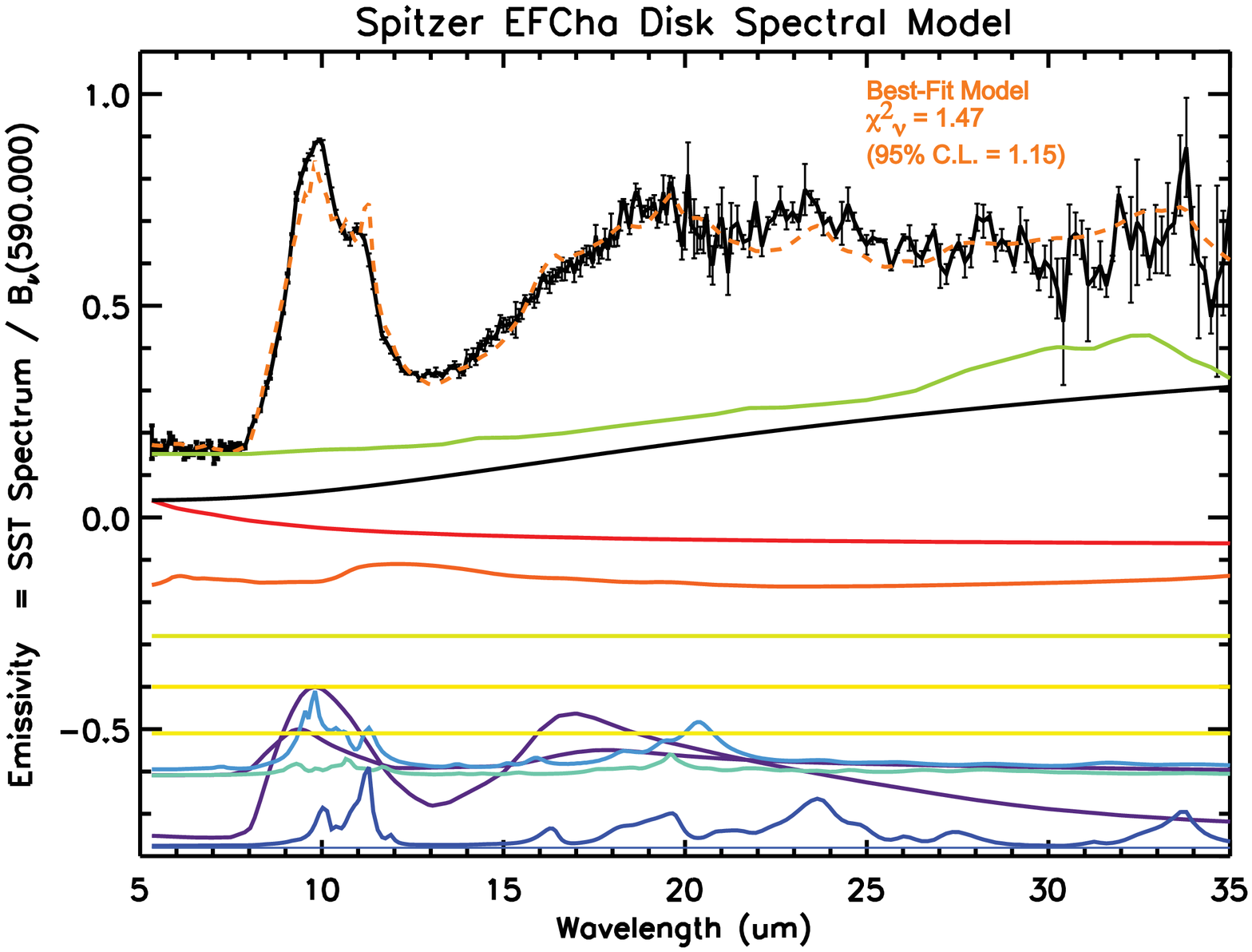}
\caption{Same as Figure \ref{efchafit} except with phyllosilicates removed from the model mix of materials and the Spitzer data re-fitted.
}
\label{efchafitnophyllo}
\end{figure}

Figure \ref{efchafitnophyllo} displays the best-fit model obtained with contributions 
from all phyllosilicate species removed.  If the best-fit model 
fares only slightly worse in reproducing the inferred emissivity with and without 
silicate contributions, then the EF Cha disk composition is underdetermined and we cannot constrain 
the existence and mass fractions of phyllosilicate species.  As shown by Figure \ref{efchafitnophyllo}, 
the $\chi^{2}$ of the best-fit model lacking phyllosilicates is substantially worse 
than for the model including phyllosilicates.
While the general shape of the emissivity is accurately 
reproduced, the phyllosilicate-free model predicts a strong peak at $\sim$ 11.25 $\mu m$, which is 
not observed, and incorrectly predicts the flux ratios of peaks at 10 $\mu m$ and 11 $\mu m$.  The model 
also significantly underpredicts the flux at 23 $\mu m$.  

\begin{figure}
\includegraphics[scale=0.6]{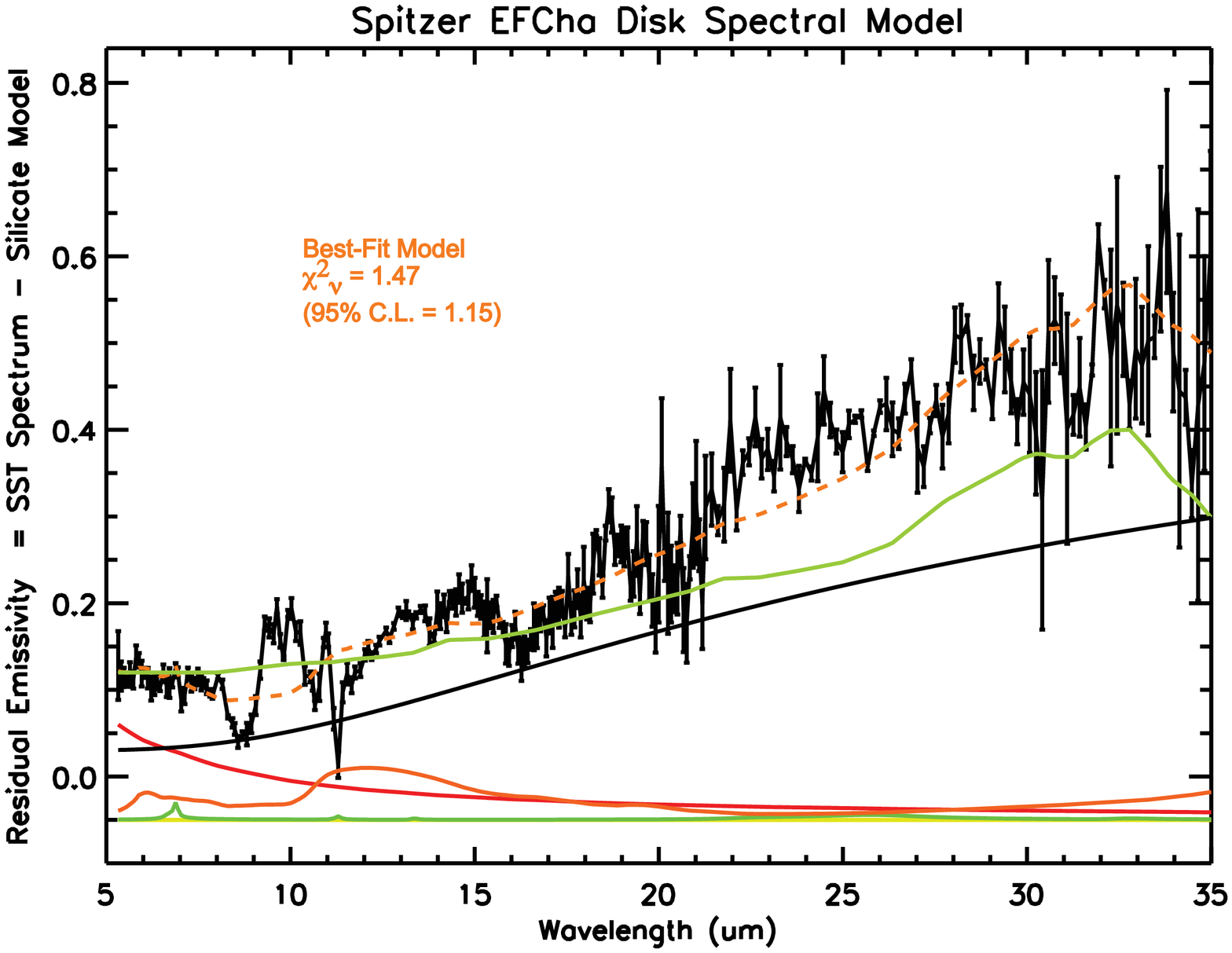}
\caption{Same as Figure \ref{efchafitnosil} except with phyllosilicates removed from the model mix of materials and the Spitzer data re-fitted.  
}
\label{efchafitnosilnophyllo}
\end{figure}

The silicate-subtracted emissivity even more strikingly reveals the failure of our phyllosilicate-free 
model (Figure \ref{efchafitnosilnophyllo}).  The model emissivity is relatively flat from 8 $\mu m$ to 11 $\mu m$.  
With the silicate contribution from the model subtracted out, the observed emissivity has local maxima 
at $\sim$ 9.5--10 $\mu m$ and 23 $\mu m$ well above that of the model.  Local minima at 8--9 $\mu m$ 
and 11-12 $\mu m$ lie well below that of the model.  The model also lacks the structure 
apparent at $\sim$ 14--20 $\mu m$ in the silicate-subtracted emissivity from the data. 
 Compared to the phyllosilicate-free model, the best-fit model including phyllosilicates 
clearly reproduces the emissivity far better.
Quantitatively, the phyllosilicate-free model can be ruled out as a 
good-fitting model.  It has a reduced $\chi^{2}$ value $\sim$ 40\% 
larger ($\chi^{2}$/dof = 1.47 for 240 degrees of freedom).  This difference in $\chi^{2}$ is statistically 
significant, since the probability that the phyllosilicate-free model fits the Spitzer data is $\sim$ 0.0002.

Table \ref{nodfit} summarizes the modeling results for our fits to the separate nod positions.  
The number of moles and weighted surface area for the species as a whole show exceptional agreement 
with the results derived for the combined spectrum.  While a few of the abundances do vary -- e.g. 
for amorphous olivine, amorphous pyroxene, smectite -- they do so by less than the 10--20\% systematic 
uncertainties on the abundances in all cases.  
While only one nod position requires the presence of a small amount of orthoenstatite, the presence/absence 
of other species is completely consistent from one nod position to the other.

Most importantly, fits to both nod positions require the presence of the three phyllosilicate species 
identified from fitting the combined spectrum.  The total weighted surface area for these species varies 
little between nod positions (0.32 vs. 0.34), and as a result the molar abundances vary little (0.19 vs. 0.20).  
Thus, our quantitative estimate of the phyllosilicate contribution to the EF Cha 
debris disk does not depend on the nod position used.  Since fitting to different nod positions probes 
the sensitivity of our results to noise characteristics, our results show that our identification 
and abundance estimate of phyllosilicate species is robust against the influence of random and systematic noise.  

At a minimum, we can conclude that some suite of species whose emissivities are extremely similar to the laboratory-measured 
emissivities for our suite of finely-powdered phyllosilicate species are abundantly present in the EF Cha debris disk. 
Given the complexity of the phyllosilicate family, and the large range of variants 
for a given phyllosilicate (e.g., the intechangability of Ca and Al, as well as Fe and Mg) 
it is not totally surprising that we may not yet have the exact laboratory match for the EF Cha spectral signature.  
Moreover, the precise shape of the spectral features depend on the exact details of 
how the minerals in the parent bodies are fragmented, combined, heated and cooled.
\textit{A priori}, we cannot rule out the existence of species not included in our chemical 
model that can mimick the emissivity of phyllosilicate species.
These uncertainties preclude any proof that the EF Cha disk must contain 
phyllosilicate material.

However, the presence of 
abundant phyllosilicate material in a variety of solar system contexts -- 
e.g. carbonaceous chondritic meteorites, Comet 9P/Tempel-1, main belt asteroids 
\citep{Tomeoka2010,Morlok2010,Lisse2006,Vilas1989,Bus2002}-- 
lends powerful support to our inference that 
the species in question is indeed phyllosilicate species.  If the species in question is 
not phyllosilicates, chemical processes must have been able to remove these species 
selectively and thoroughly, which seems implausible given the wealth of other species 
present.  Given the extensive library of species used for our spectral fitting 
and rejected as poorly matching \citep{Lisse2006}, it 
also seems unlikely that some other species is mimicking the phyllosilicate species' signatures.

\subsection{Comparisons to the Mineralogy, Dust Sizes, and Atomic Abundances of Comet Tempel-1 Ejecta and Warm, High-Luminosity Debris Disks}

To provide an additional framework for interpreting our results, here we compare 
the EF Cha dust mineralogy, dust sizes and atomic abundances to those for 
the Comet Tempel-1 ejecta and warm, young debris disks we have modeled in 
previous papers: HD 113766A and HD 172555 \citep{Lisse2006,Lisse2007,Lisse2008,Lisse2009}
  all of which demonstrate strong mid-IR features like EF Cha.
For both Comet Tempel-1 and HD 113766A, the 9--12 $\mu m$ flux exhibits a broad plateau 
with two peaks of equal amplitude at $\sim$ 10 $\mu m$ and 11 $\mu m$ due to 
abundant forsterite, enstatite, ferrosilite, and (to a lesser extent) amorphous olivine and pyroxene.    Best-fit models 
for the Comet Tempel-1 ejecta and HD 113766A also require the phyllosilicate species smectite nontronite.  
While HD 172555.  has a single sharp peak, it is at $\sim$ 9 $\mu m$ with a gradual decline in flux from 10 to 12 
$\mu m$ \citep{Lisse2009}.  Our modeling identifies this peak as due to a combination of 
amorphous silica, tektite, and SiO gas. 
EF Cha's spectrum is different from all of these, exhibiting a strong 10 $\mu m$ phyllosilicate 
peak superimposed upon a smoother 10 $\mu m$ amorphous silicate feature and 
the 9.3, 9.8, and 11.2 $\mu m$ sharp features of crystalline olivine \& pyroxene.  The phyllosilicate 
molar abundance for nontronite, talc, and saponite combined is far higher than that for HD 113766A or Comet Tempel-1 (N$_{moles, rel}$ 
= 0.20 vs. 0.03 and 0.07) and thus much more significant.  

The abundant small grain population for EF Cha is most consistent 
with a large mass fraction of debris emission being actively produced, either by a single, massive 
catastrophic collision followed by grinding of the fragments or an abnormally high, steady collision rate.  The agreement in flux density 
between the 1983 IRAS, 1996 MSX, and 2007 \textit{Spitzer} data is at least consistent with the emission being maintained 
at a high level for 20 years or more.  Timescales for grain removal by Poynting-Robertson drag 
are longer than this limit \citep[e.g.][]{BackmanParesce1993,Chen2006,CurrieKenyon2009}; however, removal timescales for the 
smallest grains due to radiation pressure are more comparable (e.g. orbital timescales of 1--100 yr), indicating that at least the smallest grains may require 
active replenishment.  Similarly, the presence of small grains and very steep particle size distribution 
is qualitatively consistent with that seen in the Tempel 1 ejecta produced by the Deep Impact collision and with
fine dust created by a hypervelocity impact found around HD 172555.

Like HD 113766A and Comet Tempel-1, the EF Cha spectrum 
may indicate the presence of water ice. HD 113766A has two debris belts beyond the ice line and 
a warm one inside of it.  Our spectral modeling suggests that EF Cha contains a warm belt well interior 
to the ice line and second one close to the ice line, at 220 K at 7.6 AU.  Water ice present in the interior 
EF Cha belt, in a radiation field 
with a local equilibrium temperature of $\sim$ 600 K, should be very quickly vaporized and thus not present 
in the mid-IR spectrum unless it has been freshly ejected from a parent body.  
Collisions producing the observed water ice then either must have occurred very recently or 
must be occurring in the outer, 220 K belt.  The belt at 7.6 AU may even provide a 
reservoir for some icy dust concentrated in the warm, terrestrial zone belt.
In either case, the presence of water ice requires an active 
replenishment source.  Moreover, if water ice is being ejected from parent bodies 
in the $\sim$ 7.6 AU region near the ice line, the EF Cha spectrum may show evidence for 
faint, water gas emission.  If water ice is indeed present there must be an active 
replenishment and a continual release of gaseous water.  We do not detect 
water gas at a statistically significant level (see Section 3.1), although 
the expected amount is difficult to quantify.  Thus, we do not rule out 
its presence at the observed signal-to-noise level.

\begin{figure}
\plotone{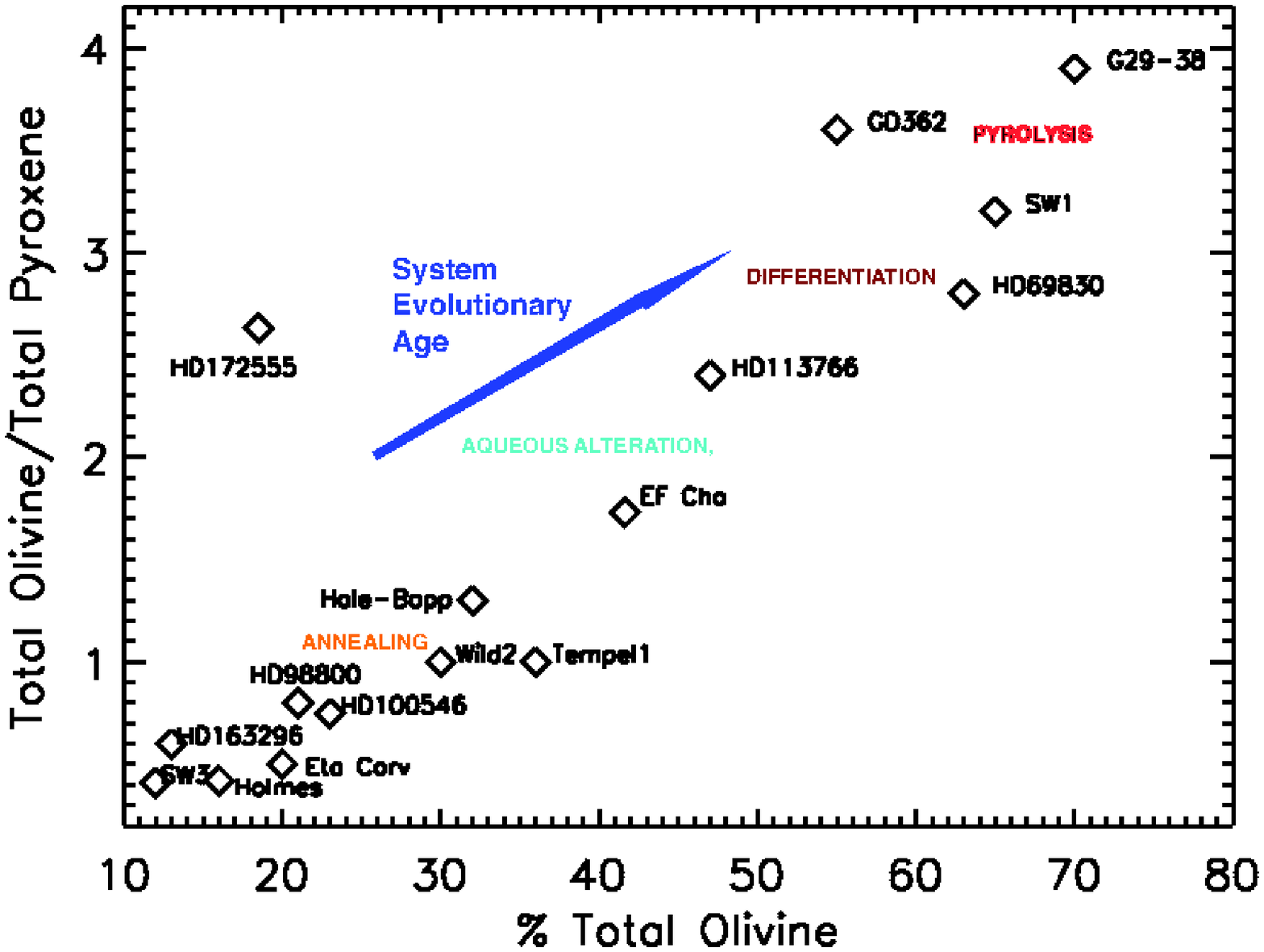}
\plotone{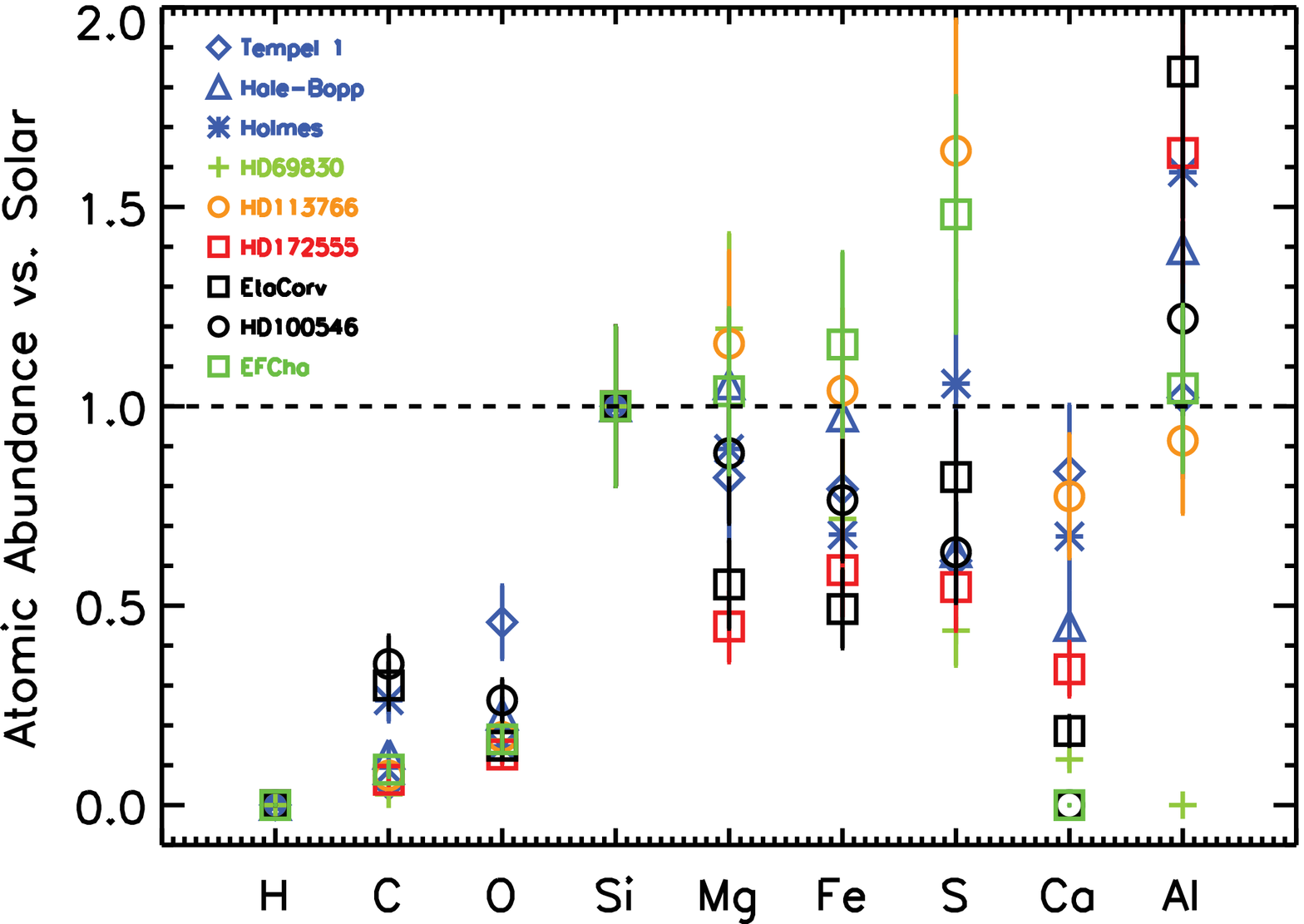}
\caption{(Top) Olivine vs. pyroxene mass fraction for EF Cha compared to that for other debris disks, protoplanetary disks, 
comets and asteroids.  The distribution of ratios define a system evolutionary sequence from unprocessed, primitive 
rock to highly processed, aqueously altered, and pyrolised rock.  (Bottom) Atomic abundances relative to solar 
for EF Cha and other debris disks, protoplanetary disks, and comets.}
\label{siltrends}
\end{figure}

Within the context established by data for comets, meteorites, protoplanetary disks, and warm debris disks, 
EF Cha's olivine mass fraction indicates that the system bridges a gap between relatively 
primitive unprocessed solids and heavily processed bodies (Figure \ref{siltrends}, top panel).  
Primitive bodies coming in contact with water, and thus aqueously altered, gain 
in their mass fractions of olivine and olivine-to-pyroxene ratios\footnote{Here we include phyllosilicates in the 
pyroxene mass fraction, assuming these species are formed by alteration of pyroxene by warm, reactive 
water \citep[see][]{Tomeoka1990,Morlok2010,Nagahara2011}.}.  Differentiated bodies and carbon-rich 
bodies processed at high temperatures have an even higher abundance of olivine and lie 
in the upper-right hand corner of the plot.  Samples with a low percentage and relative fraction of olivine occupy the lower-left 
corner and include comets (SW3, Holmes, Wild 2, and Tempel 1), and protoplanetary disks (HD 100546, HD 98800, 
HD 163296) \citep{Sitko2008,Lisse2006,Lisse2007, Lisse2008}).
  Those in the upper-right are consistent with differentiated and/or pyrolised bodies include ancient debris disks surrounding 
white dwarfs (GD-362, G29-38) and the centaur SW-1 \citep{Stansberry2004,Reach2008, Jura2006, Jura2009}. 
In this regard, EF Cha is roughly halfway between Hale-Bopp/Tempel-1 and HD 113766, which respectively identify 
the boundaries of the the primitive, unprocessed group and the highly processed, differentiated/pyrolysed group.

The bottom panel of Figure \ref{siltrends} compares atomic abundances for the EF Cha debris disk to that 
for the Sun and for other debris disks, protoplanetary disks and comets \citep[e.g. see ][]{Lisse2008}.  Like other debris/protoplanetary 
disks and comets -- HD 172555, $\eta$ Corv, HD 100546, etc -- the EF Cha emitting debris has heavily subsolar abundances 
of hydrogen, carbon and oxygen.  EF Cha has solar abundances of the main refractory elements Si, Mg, Fe, and Al, a subsolar 
abundance of calcium,  and supersolar abundances of sulphur.  Overall, EF Cha's atomic abundance pattern
shows the strongest similarity with that of HD 113766A, another luminous warm debris disk 
of comparable age whose spectrum shows evidence for massive terrestrial-zone collisions.
To first order, EF Cha's debris dust may be thought of as a more unprocessed/amorphous, water-rich analogue to 
HD 113766A's.

\subsection{Interpretation of Modeling Results}

Phyllosilicate species are tracers for the past presence of warm reactive
water, since they are produced by the aqueous alteration of anhydrous rock 
\citep[e.g.][]{Tomeoka1990}.  
Based on our modeling results, the emitting dust from EF Cha's debris disk plausibly 
includes a high abundance of phyllosilicate grains and smaller abundances of water ice and 
metal sulfides in the terrestrial zone released from parent bodies by collisions.
Because phyllosilicates should slowly decompose at temperatures of T $\sim$ 600--700 K \citep{Akai1992, Nagahara2011},
 they can exist within much of the terrestrial zone on the order of years after release from their parent body.  
However, stable low pressure phyllosilicate formation over millions of years, whether 
 from solar nebula condensation or from aqueous alteration in the interior of an asteroidal water-rich 
body, requires bodies that were formed at temperatures $\lesssim$ 300 K \citep[e.g.][]{Fegley1989,Nagahara2011}. 
Just as the presence of icy parent bodies scattered into the terrestrial zone from beyond the ice line
that can be inferred from water ice spectral features, we can conclude that the current warm dust
belt arises in part from solids imported from cold ($<$ 300K) regions, more 
distant from the star. 
The composition of the parent bodies responsible for EF Cha's debris emission is then likely not characteristic 
of planetesimals that, $\sim$ 5-10 Myr prior, grew from micron to millimeter-sized grains in the warm, terrestrial regions of 
its protoplanetary disk.  Rather, the current debris disk composition arises in part from solids imported from 
colder regions, more distant from the primary star. 

The large abundance of small grains, evidence for high collision rates, and atomic abundance pattern
 makes EF Cha's debris disk comparable to the high-luminosity, warm debris disks 
around HD 113766A.  However, its chemical composition is very different, given its 
high phyllosilicate abundance and moderately low pyroxene abundance consistent with aqueously 
altered primitive material.
If large-scale terrestrial planet formation is occuring in EF Cha like HD 113766A,
then it is occuring from a feedstock of water-rich C- or D-type asteroids, not 
water-poor S-type asteroids as inferred by Lisse et al. (2008) for HD113766A. 
  Combining our disk chemistry and grain size results, 
we conclude that the EF Cha debris disk traces 1) a high level of terrestrial 
zone collisions between bodies which, at least in part, 2) originated from beyond the 
terrestrial zone, and 3) have processed phyllosilicates from primordial silicates at some time in the past.

\section{Discussion}

\subsection{The Importance of EF Cha's Phyllosilicate Species}
The detection of abundant phyllosilicates in EF Cha’s debris disk is of particular interest
to studies of the early solar system and the habitability of terrestrial planets. Phyllosilicate
formation results from the reaction between anhydrous rock and water and thus its presence
in a debris disk requires an abundant source of water-rich planetesimals. 
Meteoritic data shows evidence for numerous "fine-grained" phyllosilicate rims surrounding 
chondrules and calcium-aluminum-rich inclusions in carbonaceous chondrites and meteorites -- i.e. 
the parent bodies that are a likely source for Earth's water \citep[e.g.][]{Robert2000} -- 
with abundances similar to phyllosilicate-rich asteroids in the outer asteroid belt (r $>$ 2.5 AU) 
\citep[]{Gradie1982,Bell1989}.  If phyllosilicate species are detected in debris disk spectra, their presence then
 requires an abundant source of water-rich planetesimals.  
Thus, the presence of phyllosilicates in the terrestrial zone of disks surrounding young stars may trace water
delivery in young planetary systems.

Moreover, because phyllosilicates trace the existence of water in disks, they may 
be important signposts for the future development of primitive life.  
Indeed, the presence of abundant phyllosilicates in the most ancient regions of Mars' surface 
has been used in part to argue that Mars may have once supported habitable conditions 
\citep{Bibring2006}.  Surface regions rich in phyllosilicates may be prime locations 
for finding early, primitive organisms \citep{Farmer1999,Orofino2010}.  

\subsection{Phyllosilicate Production and Water Delivery}
Planetsimals collisions are responsible for releasing small grains that allow us to 
identify phyllosilicate species.  However, 
it is important to emphasize that the collisions between anhydrous rock and 
hydrous planetesimals are not likely for the \textit{production} of the phyllosilicate 
species.  Impact processing tends to devolatilize planetesimals and 
should convert any phyllosilicates on the surface back to pyroxene \citep[e.g.][see section 4.3 of this work]{Morlok2010}.
Instead, phyllosilicates species are formed from the combination 
of pyroxene and warm, reactive water, where the water likely resulted from 
ice melted internal heating due to $^{26}$Al decay \citep[e.g.][, and references therein]{Desch2004,MerkPrialni2006}.   
Thus, it is more likely that the collisions are simply revealing 
aqueously altered rock as opposed to being a process that actively 
alters anhydrous rock on the planetesimals' surfaces.

Recent studies of the solar system's main belt asteroids have yielded precisely this --
detections of water and aqueously altered products \citep[e.g.][]{Campins2010,Rivkin2010}.  The total dust mass 
required to reproduce the EF Cha disk emission is comparable in mass to a 110 km-radius asteroid: a parent body of 
comparable size and density is then massive enough to have contained liquid water \citep{MerkPrialni2006}.  If our conclusion about the 
presence of phyllosilicate species is correct, it implies that asteroidal processing such as that identified from solar 
system asteroids may be quite advanced by 10 Myr.

The mechanism for the delivery of water-rich planetesimals to the terrestrial 
zone (needed for phyllosilicate production) is less unclear, although there are several plausible mechanisms.
After the planetesimal population in the outer disk is significantly depleted, the 
remaining planetesimals could be dynamically scattered inwards and incorporated into
forming terrestrial planets, especially after the solar nebula has dissipated 
\citep{Goldreich2004,Ford2007,Matsumura2010}.
Water-rich planetesimals could also be rapidly delivered from asteroid belt regions to 
the terrestrial zone by gas drag during the protoplanetary disk phase and later incorporated 
into larger bodies \citep[e.g.][]{Ciesla2005}., or form nearly 
in situ if the disk is cold.  In other systems, volatile-rich planets formed beyond the ice line 
could fail to become the cores of gas giants and migrate into the terrestrial zone through planet-disk 
or planet-planet interactions \citep{Kuchner2003}.

From the standpoint of our mineralogical analysis, all of these scenarios are equivalent. EF Cha's disk is a 
debris disk, not a protoplanetary disk.  Therefore, we do not know exactly when or how the 
water needed for phyllosilicate production was first introduced to the terrestrial zone in EF Cha.  

\subsection{The Detectability of Phyllosilicate Species in Protoplanetary Disks and Other Debris Disks}
A more answerable question is why so far only EF Cha, among all debris disks and protoplanetary disks, 
shows strong evidence for abundant phyllosilicates.  First, detecting phyllosilicate species with \textit{Spitzer} requires 
 dust warm enough to effectively emit at the 10--25 $\mu m$ with sufficiently small sizes ($<$ 10 $\mu m$) 
to show solid state features.  
Both models for terrestrial planet formation and \textit{Spitzer} surveys indicate that warm debris dust around 
young stars is rare, because the processes that can make the dust are fleeting and/or yield too few small 
collisional fragments for detectable solid state features \citep[e.g.][]{KenyonBromley2004,Chen2006,
Currie2007a,Gorlova2007}.  Thus, the subset of 
young debris disks that could possibly yield detectable mid-IR features of phyllosilicates is small 
compared to the sample of all young debris disks.

Second, experiments on phyllosilicate-rich chondritic meteorites indicate that collisions cannot 
be too high-energy, else the phyllosilicates will be destroyed.  \citet{Morlok2010} subjected 
matrix sections of the phyllosilicate-rich Murchison CM chondritic meteorite to projectile 
collisions with peak pressures ranging between 10 and 49 GPa.
The weakly shocked samples ($\sim$ 10 GPa) corresponding to low-energy collisions 
have spectra dominated by phyllosilicates, in particular 
serpentine with a broad 10 $\mu m$ peak.  For peak pressures between $\sim$ 20 and 30 GPa, 
the Murchison samples retain mid-IR signatures of phyllosilicate species but crystalline olivine's 
peak at $\sim$ 11.2 $\mu m$ begins to emerge.  At higher pressures ($\sim$ 36 GPa) corresponding 
to high-energy collisions, the phyllosilicate 
signatures are lost due to serpentine decomposition and devolatization and the spectra better resemble 
those for debris disks with weaker/absent phyllosilicate emission such as HD 113766A and HD 69830.
Thus, EF Cha's debris disk plausibly represents a particularly rare laboratory for studying 
phyllosilicates in other planetary systems, where collisions are sufficient to produce detectable 
debris but not so catastrophic as to eliminate evidence for aqueously altered planetesimals.

While numerous protoplanetary disks in nearby star-forming regions have been targeted by 
Spitzer, studies typically do not incorporate phyllosilicate species in their spectral 
modeling and instead focus on amorphous silicates, forsterite, 
enstatite, and silica \citep[e.g.][]{Sargent2009,Watson2009,Juhasz2010}
\footnote{Some of these studies identify solid state features in many 
protoplanetary disks at wavelengths qualitatively similar to phyllosilicate 
features.  However, these authors argue that the features are due 
to allotropes of silica \citep[e.g.][]{Sargent2009, Juhasz2010}, which we have been able to 
rule out for EF Cha.}.
Recent models for protoplanetary disk spectra by \citet{MorrisDesch2010} 
argue that phyllosilicates can be identified at $\sim$ 10 $\mu m$ and, 
especially, 20--25 $\mu m$ if they comprise more than 3\% of the dust by surface 
area, as is the case for EF Cha, and if the dust is fine enough to be optically thin
 \footnote{Their modeling formalism differs from ours in that they adopt a ``distribution of 
hollow spheres" to compute dust opacities.   We use laboratory emission spectra from 
fine powders.  However, their computed emissivities qualititatively capture 
basic features of those for powders, specifically a strong, narrow peak at 10 $\mu m$.}.

However, our analysis cannot yet constrain the frequency or duration of phyllosilicate production in debris disks.
No other debris disk has been reported to date as having strong, clearly phyllosilicate derived emission features.
Further, since debris disk grains are generally too large to produce clear solid-state features anyway, 
we cannot conclude that phyllosilicate production in debris disks is fleeting or rare.  
More precisely, determining exactly when and how often phyllosilicates form, and thus when water is delivered 
to the terrestrial zone of young planetary systems, will require modeling more protoplanetary disks and 
debris disks with solid state features using sophisticated mineralogical models like those 
presented here.   
We are encouraged that future work will produce more and more phyllosilicate detections, 
however, given the preponderance of likely phyylosilicate rich C- and D-type asteroids 
in the outer main belt of the solar system, and the suggestion of phyllosilicates in 
comets (Lisse et al. 2006, 2007a, Reach et al. 2009a).

Spectral modeling of protoplanetary disks is also arguably more complicated, given 
the effects of disk flaring/shadowing, variability, and the inability to 
probe the disk midplane (where planetesimals should be located) if the disk 
is optically thick.  However, many Herbig AeBe stars with protoplanetary disks
in the \citeauthor{Juhasz2010} study (HD 36112, HD 37258, HD 37357, etc.) --
-- the likely evolutionary precursors to older A stars with debris disks 
like EF Cha -- have high signal-to-noise \textit{Spitzer} spectra and numerous 
solid state features in the 10--25 $\mu m$ range.  While many (most?) of their disks 
can be fit by typical olivine+pyroxene+silica porous Mie spectral models and PAH empirical 
residual templates \citep[e.g.][]{Juhasz2010},  
they also provide an interesting future sample to apply our spectral modeling to 
search for phyllosilicate-rich protoplanetary disks.  

\section{Conclusion}
This paper presents analysis of \textit{Spitzer} IRS data for the debris disk surrounding the 10 Myr-old star EF Cha.  
To investigate the composition of the EF Cha debris disk, location of the debris dust with respect to the star, and 
size of the emitting grains, we have compared our disk spectra to that for disks with a 
wide range of evolutionary stages and analyze it using a sophisticated mineralogical 
model.  Our study yields the following main results:

\begin{itemize}

\item Using simple energy balance considerations, the EF Cha disk must be a gas-poor debris disk with LTE dust temperatures between 
225 K and 430 K and dust physically located at $\sim$ 1--7.5 AU.  Because 
these temperatures are characteristic of equilibrium dust temperatures between 0.45 and 1.55 AU of the 
Sun in our solar system, we expect that the EF Cha debris disk contains terrestrial zone material. 

\item Unlike most debris disks, the EF Cha disk shows evidence for numerous solid state 
features diagnostic of the composition of parent bodies producing its debris, implying 
consistent presence of small, $\mu m$-sized grains around the star.
Comparing the EF Cha spectrum with that for other disks exhibiting solid state features (mostly 
protoplanetary disks), shows that the EF Cha disk lacks evidence for PAH emission and has 
relatively little crystalline olivine and pyroxene.
  While the general shape of the EF Cha spectrum is 
consistent with a high abundance of amorphous silicates, numerous maxima secondary features
 show that additional species must be abundantly present.  

\item  Minerological modeling shows that the EF Cha debris disk includes an exceptionally high 
mass fraction of phyllosilicates, produced by aqueous alteration of anhydrous rock,
making it unique amongst all debris disks studied thus far.  
Its elevated olivine to pyroxene ratio is further evidence of large-scale aqueous alteration.
The dust may also contains water ice and metal sulfides.

\item The amount of dust present is the equivalent of a large, $\sim$ 110 
km radius asteroid, implying the participation of a large parent body 
or bodies sourcing the observed dust; this sourcing could have been caused by collisions in an asteroid belt. 

\item Because only a narrow range of collisional energieswill yield detectable debris but not 
convert phyllosilicates back into silicates via devolitalization, 
EF Cha's debris disk represents a potentially rare laboratory for studying 
phyllosilicate species -- a signpost of terrestrial water delivery -- in a debris disk and 
motivates future studies to identify these species in protoplanetary disks.

\end{itemize}

\acknowledgements
We thank Marc Kuchner, Aki Roberge, Scott Kenyon, Alycia Weinberger, John Debes, 
Steve Desch, Melissa Morris, and the anonymous referee for useful comments and discussions 
that improved the quality of this work.
This paper was based on observations taken with the NASA \textit{Spitzer Space Telescope}, 
operated by JPL/CalTech. T. Currie is funded by a NASA Postdoctoral Fellowship; C. M. Lisse 
gratefully acknowledges support for performing the modeling described herein from 
JPL contract 1274485, NSF AAG grant NNX09AU31G-1, and the APL Janney Fellowship program. 
A. Sicilia-Aguilar acknowledges support from the Deutsche Forschungsgemeinschaft (DFG) 
grant SI 1486/1-1.  K. Y. L. Su and G. Rieke were partially supported by contract 
1255094 from Caltech/JPL to the University of Arizona.

{}
\begin{deluxetable}{llllllllllllll}
 \tiny
\setlength{\tabcolsep}{0.1in}
\tabletypesize{\tiny}
\tablecolumns{11}
\tablecaption{Composition of the Best-Fit Model to the Combined \textit{Spitzer} IRS EF Cha Spectrum}
\tiny
\tablehead{{Species}&{Surface Area}&{Density}&{M.W.} & {N$_{moles}$}& {N$_{moles}$ (range)}&{Mass}&{Model T$_{Max}$}&{Model $\chi^{2}$} \\
{}&{(Weighted)}&{g cm$^{-3}$}&{}&{(rel.)}&{rel.}&{(rel.)}&{(K)}&{if not included}}
\startdata
\textbf{\textit{Olivines}}&&&&&&\\
 Amorph. Olivine (MgFeSiO$_{4}$)&0.35&3.6 &172&0.73 & [0.66,0.80]&125.5 &570&35.2\\   
 Forsterite (Mg$_{2}$SiO$_{4}$) & 0.20 & 3.2 & 140 & 0.46 & [0.41,0.51]&64.4& 570 & 2.54\\
 Fayalite (Fe$_{2}$SiO$_{4}$)  & 0.02 & 4.3 & 204 & 0.04 & [0.02,0.06]&8.1 & 570 & 1.07\\
\\
\textbf{\textit{Pyroxenes}}\\
 Amorph. Pyroxene(MgFeSi$_{2}$O$_{6}$)& 0.06 & 3.5 &232 & 0.09 &[0.08,0.10]&20.9& 570 & 1.54\\
 FerroSilite(Fe$_{2}$Si$_{2}$O$_{6}$) & 0.00 & 4.0 & 264 & 0.00 & [0,0]&0&550 & 1.04\\
 Diopside (CaMgSi$_{2}$O$_{6}$) & 0.00 & 3.3 & 216 & 0.00 & [0,0]&0&570 & 1.04\\
 OrthoEnstatite (Mg$_{2}$Si$_{2}$O$_{6}$) & 0.02 & 3.2 & 200 & 0.03 & [0,0.04]&6.0& 570 & 1.07\\
\\
\textbf{\textit{Phyllosilicate Species}}\\
 Smectite (Nontronite)& 0.10 & 2.3 & 496 & 0.05 & [0.04,0.06]&24.8&570 & 3.10\\
 Talc (Mg$_{3}$Si$_{4}$O$_{10}$(OH)$_{2}$) & 0.16 & 2.8 & 379 & 0.12 & [0.10,0.13]&45.5 &570 & 1.5\\
 Saponite & 0.06 & 2.3 & 480 & 0.03 & [0.02,0.03]&14.4 & 570 & 1.27\\
\\
\textbf{\textit{Carbonates}}\\
  Magnesite (MgCO$_{3}$) & 0.02 & 3.1 & 84 & 0.07 & [0.06,0.08]&5.88&570 & 1.09\\
  Siderite (FeCO$_{3}$) & 0.00 & 3.9 & 116 & 0.00 & [0,0]&0&570 & 1.04\\
\\
\textbf{\textit{Metal Sulfides}}\\
 Niningerite (Mg$_{10}$Fe$_{90}$S)  & 0.27 & 4.5 & 84 & 1.45 & [1.31,1.60]&121.8&570 & 7.82\\            
\\
\textbf{\textit{Water}}\\
 Water Ice & 0.08 & 1 & 18  & 0.44 & [0.40,0.49]&7.92&190 & 4.66\\
 Water Gas & 0.00 & 1 & 18 & 0.00 & [0,0]&0&190 & 1.04\\
\\
\textbf{\textit{Organics}}
\\
 Amorph. Carbon & 0.09 & 2.5 & 12 & 1.88 & [1.69,2.07]&22.6&715 & 6.73\\
 PAH (C$_{10}$H$_{14}$)  & 0.00 & 1.0 & 178 & 0.00 & [0,0]&0&N/A & 1.04
\enddata
\tablecomments{The relative number of moles is calculated as follows: 
N$_{moles}$(i) $\sim$ Density(i)/Molecular Weight(i)$\times$Surface Area$_{weighted}$.  
The relative masses of each species are calculated from the product of N$_{moles}$ and 
the molecular weight.  N$_{moles}$ (range) identifies the range of N$_{moles}$ from 
models that quantitatively provide a good fit to the data ($\chi^{2}$ $<$ $\chi^{2}$ (95\% C.L.)).
}

\end{deluxetable}

\begin{deluxetable}{llllllllllllll}
 \tiny
\setlength{\tabcolsep}{0.1in}
\tabletypesize{\tiny}
\tablecolumns{11}
\tablecaption{Composition of the Best-Fit Model to the \textit{Spitzer} IRS EF Cha Spectrum for
Different Nod Positions}

\tiny
\tablehead{{Species}&{Surface Area}&{Density}&{M.W.} & {N$_{moles}$}& {Model T$_{Max}$} \\
{}&{(Weighted)}&{g cm$^{-3}$}&{}&{(rel.)}&{(K)}}
\startdata
\textbf{\textit{Olivines}}&&&&&&\\
 Amorph. Olivine (MgFeSiO$_{4}$)&0.37/0.35&3.6 &172&0.77/0.73& 570\\   
 Forsterite (Mg$_{2}$SiO$_{4}$) & 0.20/0.20 & 3.2 & 140 & 0.46/0.46 & 570\\
 Fayalite (Fe$_{2}$SiO$_{4}$)  & 0.01/0.03 & 4.3 & 204 & 0.02/0.06 & 570\\
\\
\textbf{\textit{Pyroxenes}}\\
 Amorph. Pyroxene(MgFeSi$_{2}$O$_{6}$)& 0.06/0.05 & 3.5 &232 & 0.09/0.08 & 570 \\
 FerroSilite(Fe$_{2}$Si$_{2}$O$_{6}$) & 0.00/0.00 & 4.0 & 264 & 0.00/0.00 & 550\\
 Diopside (CaMgSi$_{2}$O$_{6}$) & 0.00/0.00 & 3.3 & 216 & 0.00/0.00 & 570\\
 OrthoEnstatite (Mg$_{2}$Si$_{2}$O$_{6}$) & 0.00/0.02 & 3.2 & 200 & 0.00/0.03 & 570\\
\\
\textbf{\textit{Phyllosilicate Species}}\\
 Smectite (Nontronite)& 0.09/0.11 & 2.3 & 496 & 0.04/0.05 & 570 \\
 Talc (Mg$_{3}$Si$_{4}$O$_{10}$(OH)$_{2}$) & 0.16/0.16 & 2.8 & 379 & 0.12/0.12 & 570 \\
 Saponite & 0.07/0.07 & 2.3 & 480 & 0.03/0.03 & 570 \\
\\
\textbf{\textit{Carbonates}}\\
  Magnesite (MgCO$_{3}$) & 0.01/0.01 & 3.1 & 84 & 0.07/0.07 & 570 \\
  Siderite (FeCO$_{3}$) & 0.00/0.00 & 3.9 & 116 & 0.00/0.00 & 570 \\
\\
\textbf{\textit{Metal Sulfides}}\\
 Niningerite (Mg$_{10}$Fe$_{90}$S)  & 0.28/0.27 & 4.5 & 84 & 1.5/1.45 & 570 \\            
\\
\textbf{\textit{Water}}\\
 Water Ice & 0.08/0.08 & 1 & 18  & 0.44/0.44 & 190 \\
 Water Gas & 0.00/0.00 & 1 & 18 & 0.00/0.00 & 190 \\
\\
\textbf{\textit{Organics}}
\\
 Amorph. Carbon & 0.09/0.09 & 2.5 & 12 & 1.88/1.88 & 715\\
 PAH (C$_{10}$H$_{14}$)  & 0.00/0.00 & 1.0 & 178 & 0.00/0.00 & N/A 
\enddata
\tablecomments{The two entries for the weighted surface area and number of moles
 for the first nod position and then the second nod position.
 With the exception of orthoenstatite, our identification of species from modeling the
combined spectrum are not affected by the noise characteristics of the spectra.}
\label{nodfit}

\end{deluxetable}

\clearpage



\end{document}